\documentclass[aps,prd,twocolumn,superscriptaddress,nofootinbib]{revtex4-1}

\usepackage{latexsym}
\usepackage{amsmath}
\usepackage{amssymb}
\usepackage{amsfonts}
\usepackage{bm}

\usepackage{color}
\definecolor{purple}{rgb}{0.5,0,0.5}
\definecolor{blue}{rgb}{0.0,0,0.9}
\definecolor{prdblue}{rgb}{0.133,0.118,0.498}
\usepackage[colorlinks=true, pdfstartview=FitV, linkcolor=prdblue, citecolor= prdblue, urlcolor=prdblue]{hyperref}

\usepackage{supertabular} 
\usepackage{placeins}
\usepackage{epsfig}
\usepackage{graphicx}

\begin{document}


\title{Hidden-charm pentaquarks with strangeness in a chiral quark model}


\author{Gang Yang}
\email[]{yanggang@zjnu.edu.cn}
\affiliation{Department of Physics, Zhejiang Normal University, Jinhua 321004, China}

\author{Jialun Ping}
\email[]{jlping@njnu.edu.cn}
\affiliation{Department of Physics and Jiangsu Key Laboratory for Numerical Simulation of Large Scale Complex Systems, Nanjing Normal University, Nanjing 210023, P. R. China}

\author{Jorge Segovia}
\email[]{jsegovia@upo.es}
\affiliation{Departamento de Sistemas F\'isicos, Qu\'imicos y Naturales, Universidad Pablo de Olavide, E-41013 Sevilla, Spain}


\date{\today}

\begin{abstract}
The LHCb collaboration has recently announced the discovery of two hidden-charm pentaquark states with also strange quark content, $P_{cs}(4338)$ and $P_{cs}(4459)$; its analysis points towards having both hadrons isospin equal to zero and spin-parity quantum numbers $\frac12^-$ and $\frac32^-$, respectively. We perform herein a systematical investigation of the $qqsc\bar{c}$ $(q=u,\,d)$ system by means of a chiral quark model, along with a highly accurate computational method, the Gaussian expansion approach combined with the complex-scaling technique. Baryon-meson configurations in both singlet- and hidden-color channels are considered. The $P_{cs}(4338)$ and $P_{cs}(4459)$ signals can be well identified as molecular bound states with dominant components $\Lambda J/\psi$ $(60\%)$ and $\Xi_c D$ $(23\%)$ for the lowest-energy case and $\Xi_c D^*$ $(72\%)$ for the highest-energy one. Besides, it seems that some narrow resonances can be also found in each allowed $I(J^P)$-channel in the energy region of $4.6-5.5$ GeV, except for the $1(\frac12^-)$ where a shallow bound state with dominant $\Xi^*_c D^*$ structure is obtained at $4673$ MeV with binding energy $E_B=-3$ MeV. These exotic states are expected to be confirmed in future high energy experiments.
\end{abstract}


\maketitle


\section{INTRODUCTION}
\label{sec:Introduction}

During the past few years, some hadrons with $5$-quark content (pentaquarks) have been reported experimentally. In particular, the hidden-charm pentaquark candidate $P_c(4380)^+$ was the first to announce by the LHCb collaboration in 2015~\cite{Aaij:2015tga}. After then, three more pentaquark states, with equal minimum quark content of $qqqc\bar{c}$ $(q=u,\,d)$, $P_c(4312)^+$, $P_c(4440)^+$ and $P_c(4457)^+$ were reported by the same collaboration in 2019~\cite{lhcb:2019pc}. Additionally, progress in hidden-charm pentaquarks with strangeness has also been made by the LHCb collaboration since 2020. By using $pp$ collision data, a $\Lambda J/\psi$ structure, which is labeled as the $P_{cs}(4459)$, was reported in $\Xi^-_b \rightarrow J/\psi \Lambda K^-$ decays~\cite{LHCb:2020jpq}. Mass and width of this hidden-charm pentaquark candidate with strange quark content are $4458.8\pm 2.9^{+4.7}_{-1.1}$ MeV and $17.3\pm 6.5^{+8.0}_{-5.7}$ MeV, respectively. In 2022, another strange pentaquark candidate, $P_{cs}(4338)$, was observed with high significance in $B^- \rightarrow J/\psi \Lambda \bar{p}$ decays~\cite{LHCb:2022ogu}. Its experimental mass and width are $4338.2\pm 0.7 \pm 0.4$ MeV and $7.0\pm 1.2 \pm 1.3$ MeV. The Spin-parity of these two exotic states, $P_{cs}(4338)$ and $P_{cs}(4459)$, are preferably to be $\frac12^-$ and $\frac32^-$, repectively.

These facts have triggered an enormous amount of theoretical investigations; concerning the hidden-charm pentaquark candidates with strangeness one may mention, for instance, the works done within effective field theories~\cite{Feijoo:2022rxf, Zhu:2022wpi, Yan:2022wuz, Chen:2022onm, Du:2021bgb, Zhu:2021lhd}, using QCD sum rules~\cite{Wang:2022neq, Wang:2022gfb, Wang:2020eep, Chen:2020uif, Azizi:2023foj} or based on phenomenological quark models~\cite{Ortega:2022uyu, Giachino:2022pws, Yang:2022ezl, Wang:2022mxy, Karliner:2022erb, Meng:2022wgl, Shi:2021wyt, Xiao:2021rgp}, generally stablishing that the $P_{cs}(4338)$ state can be identified as a $\Xi_c \bar{D}^{(*)}$ molecular structure whereas the $P_{cs}(4459)$ seems better to be a $\Xi^{(',*)}_c \bar{D}^{(*)}$ hadronic molecule. However, mixed configurations~\cite{Chen:2020kco, Chen:2022wkh}, compact structure analysis~\cite{Li:2023aui, Maiani:2023nwj} and triangle singularities~\cite{Burns:2022uha} could also explain the nature of the mentioned $P_{cs}$ states. In addition, several additional exotic states in the hidden-charm pentaquark sector with strange quark content are theoretically proposed in Refs.~\cite{Ke:2023nra, Yalikun:2023waw, Wu:2021gyn, Wang:2023iox}. Besides, the electromagnetic properties of the mentioned pentaquarks are calculated in Refs.~\cite{Ozdem:2023htj, Wang:2022tib, Ozdem:2022kei, Li:2021ryu}. The production and decay properties are also studied in Refs.~\cite{Paryev:2023icm, Cheng:2021gca, Yang:2021pio, Chen:2021tip, Azizi:2021utt, Liu:2020ajv, Wu:2021caw, Lu:2021irg}.

Within a chiral quark model approach~\cite{Vijande:2004he, Segovia:2013wma}, supplemented by employing a highly accurate computational method on few-body problems, \emph{i.e.} the combination of Gaussian expansion method (GEM)~\cite{Hiyama:2003cu} and the complex-scaling method (CSM)~\cite{Myo:2014ypa}, the S-wave hidden-charm pentaquarks with strangeness, and having spin-parity $\frac12^-$, $\frac32^-$ and $\frac52^-$, in the iso-scalar and -vector sectors, are systematically investigated. This theoretical framework has already been successfully applied in various multiquark systems: charmonium- and bottomonium-like tetraquarks~\cite{Yang:2023mov, Yang:2021zhe, Yang:2022cut}; singly, doubly and fully heavy tetraquarks~\cite{Yang:2021hrb, Yang:2019itm, Yang:2020fou, Yang:2021izl}; and hidden-charm and -bottom, doubly and fully heavy pentaquarks~\cite{Yang:2015bmv, Yang:2018oqd, Yang:2020twg, Yang:2022bfu}. Therein, many of the experimentally announced exotic states, such as the $X(2900)$, $X(6900)$, $X(3985)$, $P_c(4312)$ and $P_c(4380)$, may be well identified. Therefore, it is also an natural extension of our theoretical investigation to incorporate the analysis of pentaquark systems with strange quark content, beginning with the hidden-charm pentaquarks with strangeness motivated by the recently reported $P_{cs}$ signals.

We arrange the present work as following parts. In Sec.~\ref{sec:model} the chiral quark model, pentaquark wave-functions, GEM and CSM are briefly presented and discussed. Section~\ref{sec:results} is devoted to the analysis and discussion on the obtained results. Finally, a summary is presented in Sec.~\ref{sec:summary}.


\section{THEORETICAL FRAMEWORK}
\label{sec:model}

A throughout review of the theoretical formalism can be found in Ref.~\cite{Yang:2020atz}. We shall then focus here on the most relevant features of the phenomenological model and the numerical method concerning the hidden-charm pentaquarks with strangeness $qqsc\bar{c}$ $(q=u,\,d)$.

\subsection{The Hamiltonian}

The general form of the non-relativistic five-body Hamiltonian in a complex scaling method can be written as
\begin{equation}
H(\theta) = \sum_{i=1}^{5}\left( m_i+\frac{\vec{p\,}^2_i}{2m_i}\right) - T_{\text{CM}} + \sum_{j>i=1}^{5} V(\vec{r}_{ij} e^{i\theta}) \,,
\label{eq:Hamiltonian}
\end{equation}
In Eq.~(\ref{eq:Hamiltonian}), $T_{\text{CM}}$ is the center-of-mass kinetic energy and the two-body potential,
\begin{equation}\label{CQMV}
V(\vec{r}_{ij} e^{i\theta}) = V_{\text{CON}}(\vec{r}_{ij} e^{i\theta}) + V_{\text{OGE}}(\vec{r}_{ij} e^{i\theta}) + V_{\chi}(\vec{r}_{ij} e^{i\theta}) \,,
\end{equation}
includes color-confining, one-gluon-exchange and Goldstone-boson exchange interactions. Besides, the coordinates that describe relative motion between quarks are transformed with a complex rotation, $\vec{r} \rightarrow \vec{r} e^{i\theta}$. Accordingly, within the framework of complex-range, the dynamics of a five-body system is determined by solving a complex scaled Schr\"{o}dinger equation:
\begin{equation}\label{CSMSE}
\left[ H(\theta)-E(\theta) \right] \Psi_{JM_J}(\theta)=0 \,,
\end{equation}
where the (complex) eigenvalue $E$ can be assigned into three types of singularities: bound, resonance and scattering states. Particularly, bound and resonance states are independent of the rotated angle $\theta$, with the first one placed on the real-axis of a complex energy plane, and the second one located above the continuum threshold with a total strong decay width $\Gamma=-2\,\text{Im}(E)$. On the other hand, the scattering states depend on the rotated angle and follows the path marked by its associated continuum threshold.

Some details about the different potential terms in Eq.~\eqref{CQMV} and its physical motivation come now. Firstly, color confinement should be encoded in the non-Abelian character of Quantum Chromodynamics (QCD). Some studies of QCD on a lattice have demonstrated that multi-gluon exchanges produce an attractive linearly rising potential proportional to the distance between infinite-heavy quarks~\cite{Bali:2005fu}. However, the spontaneous creation of light-quark pairs from the QCD vacuum may give rise at the same scale to a breakup of the color flux-tube~\cite{Bali:2005fu}. Our potential model tries to mimic these two phenomenological observations by the following expression, in complex scaling method,
\begin{equation}
V_{\text{CON}}(\vec{r}_{ij} e^{i\theta}\,)=\left[-a_{c}(1-e^{-\mu_{c}r_{ij} e^{i\theta}})+\Delta \right] 
(\vec{\lambda}_{i}^{c}\cdot \vec{\lambda}_{j}^{c}) \,,
\label{eq:conf}
\end{equation}
where $a_{c}$ and $\mu_{c}$ are model parameters, and $\vec{\lambda}^c$ denotes the SU(3) color Gell-Mann matrix. The potential of Eq.~\eqref{eq:conf} is linear at short inter-quark distances with an effective confinement strength $\sigma = -a_{c} \, \mu_{c} \, (\vec{\lambda}^{c}_{i}\cdot \vec{\lambda}^{c}_{j})$, while it becomes constant at large distances, $V_{\text{thr}} = (\Delta-a_c)(\vec{\lambda}^{c}_{i}\cdot \vec{\lambda}^{c}_{j})$. 

The QCD perturbative effects are taken into account through one-gluon fluctuations around the instanton vacuum. Then, the different terms of the potential derived from the Lagrangian,
\begin{eqnarray}
{\mathcal L}_{qqg} &=& i\sqrt{4\pi\alpha_s} \bar \psi \gamma_\mu G^\mu_c
\lambda^c \psi,
\label{Lqqg}
\end{eqnarray}
contain central, tensor and spin-orbit contributions. For a $S$-wave pentaquark system, we consider herein only the central one, expressed also with a complex transformation $\vec{r} \rightarrow \vec{r} e^{i\theta}$, 
\begin{align}
V_{\text{OGE}}(\vec{r}_{ij} e^{i\theta}) &= \frac{1}{4} \alpha_{s} (\vec{\lambda}_{i}^{c}\cdot
\vec{\lambda}_{j}^{c}) \Bigg[\frac{1}{r_{ij} e^{i\theta}} \nonumber \\ 
&
- \frac{1}{6m_{i}m_{j}} (\vec{\sigma}_{i}\cdot\vec{\sigma}_{j}) 
\frac{e^{-r_{ij} e^{i\theta}/r_{0}(\mu)}}{r_{ij} e^{i\theta} r_{0}^{2}(\mu)} \Bigg] \,,
\end{align}
where $m_{i}$ is the quark mass and $\vec{\sigma}$ denotes the Pauli matrices. The contact term has been regularized as follows
\begin{equation}
\delta(\vec{r}_{ij} e^{i\theta})\sim\frac{1}{4\pi r_{0}^{2}}\frac{e^{-r_{ij} e^{i\theta}/r_{0}}}{r_{ij} e^{i\theta}} \,,
\end{equation}
with $r_{0}(\mu_{ij})=\hat{r}_{0}/\mu_{ij}$ depending on $\mu_{ij}$, the reduced mass of a quark--(anti-)quark pair.

The wide energy range needed to provide a consistent description of mesons and baryons, from light to heavy quark sectors, requires an effective scale-dependent strong coupling constant~\cite{Segovia:2013wma}
\begin{equation}
\alpha_{s}(\mu_{ij})=\frac{\alpha_{0}}{\ln\left(\frac{\mu_{ij}^{2}+\mu_{0}^{2}}{\Lambda_{0}^{2}} \right)} \,,
\end{equation}
where $\alpha_{0}$, $\mu_{0}$ and $\Lambda_{0}$ are model parameters.

Dynamical chiral symmetry breaking is the mechanism responsible for making light quarks, with very small current masses, acquire a dynamical, momentum dependent mass $M(p)$, with $M(0)\approx 300\,\mbox{MeV}$ for the $u$ and $d$ quarks, namely the constituent quark mass. To preserve chiral invariance of the QCD Lagrangian new interaction terms, given by Goldstone boson exchanges, must appear. The central terms of the chiral quark--(anti-)quark interaction $V_{\chi}(\vec{r}_{ij} e^{i\theta})$ can be written as the following four parts,
\begin{align}
&
V_{\pi}\left( \vec{r}_{ij} e^{i\theta} \right) = \frac{g_{ch}^{2}}{4\pi}
\frac{m_{\pi}^2}{12m_{i}m_{j}} \frac{\Lambda_{\pi}^{2}}{\Lambda_{\pi}^{2}-m_{\pi}
^{2}}m_{\pi} \Bigg[ Y(m_{\pi}r_{ij} e^{i\theta}) \nonumber \\
&
\hspace*{1.20cm} - \frac{\Lambda_{\pi}^{3}}{m_{\pi}^{3}}
Y(\Lambda_{\pi}r_{ij} e^{i\theta}) \bigg] (\vec{\sigma}_{i}\cdot\vec{\sigma}_{j})\sum_{a=1}^{3}(\vec{\lambda}_{i}^{a}
\cdot \vec{\lambda}_{j}^{a}) \,, \\
& 
V_{\sigma}\left( \vec{r}_{ij} e^{i\theta} \right) = - \frac{g_{ch}^{2}}{4\pi}
\frac{\Lambda_{\sigma}^{2}}{\Lambda_{\sigma}^{2}-m_{\sigma}^{2}}m_{\sigma} \Bigg[
Y(m_{\sigma}r_{ij} e^{i\theta}) \nonumber \\
&
\hspace*{1.20cm} - \frac{\Lambda_{\sigma}}{m_{\sigma}}Y(\Lambda_{\sigma}r_{ij} e^{i\theta})
\Bigg] \,, \\
& 
V_{K}\left( \vec{r}_{ij} e^{i\theta} \right)= \frac{g_{ch}^{2}}{4\pi}
\frac{m_{K}^2}{12m_{i}m_{j}} \frac{\Lambda_{K}^{2}}{\Lambda_{K}^{2}-m_{K}^{2}}m_{
K} \Bigg[ Y(m_{K}r_{ij} e^{i\theta}) \nonumber \\
&
\hspace*{1.20cm} -\frac{\Lambda_{K}^{3}}{m_{K}^{3}}Y(\Lambda_{K}r_{ij} e^{i\theta})
\Bigg] (\vec{\sigma}_{i}\cdot\vec{\sigma}_{j})\sum_{a=4}^{7}(\vec{\lambda}_{i}^{a}
\cdot \vec{\lambda}_{j}^{a}) \,, \\
& 
V_{\eta}\left( \vec{r}_{ij} e^{i\theta} \right) = \frac{g_{ch}^{2}}{4\pi}
\frac{m_{\eta}^2}{12m_{i}m_{j}} \frac{\Lambda_{\eta}^{2}}{\Lambda_{\eta}^{2}-m_{
\eta}^{2}}m_{\eta} \Bigg[ Y(m_{\eta}r_{ij} e^{i\theta}) \nonumber \\
&
\hspace*{1.20cm} -\frac{\Lambda_{\eta}^{3}}{m_{\eta}^{3}
}Y(\Lambda_{\eta}r_{ij} e^{i\theta}) \Bigg] (\vec{\sigma}_{i}\cdot\vec{\sigma}_{j})
\Big[\cos\theta_{p} \left(\vec{\lambda}_{i}^{8}\cdot \vec{\lambda}_{j}^{8}
\right) \nonumber \\
&
\hspace*{1.20cm} -\sin\theta_{p} \Big] \,,
\end{align}
where $Y(x)$ is the Yukawa function defined by $Y(x)=e^{-x}/x$. The physical $\eta$ meson is considered by introducing the angle $\theta_p$. The $\vec{\lambda}^{a}$ is the SU(3) flavor Gell-Mann matrix. Taken from their experimental values, $m_{\pi}$, $m_{K}$ and $m_{\eta}$ are the masses of the SU(3) Goldstone bosons. The value of $m_{\sigma}$ is determined through the partially conserved axial current (PCAC) relation $m_{\sigma}^{2}\simeq m_{\pi}^{2}+4m_{u,d}^{2}$~\cite{Scadron:1982eg}. Finally, the chiral coupling constant, $g_{ch}$, is determined from the $\pi NN$ coupling constant through
\begin{equation}
\frac{g_{ch}^{2}}{4\pi}=\frac{9}{25}\frac{g_{\pi NN}^{2}}{4\pi} \frac{m_{u,d}^{2}}{m_{N}^2} \,,
\end{equation}
which assumes that flavor SU(3) is an exact symmetry only broken by the different mass of the strange quark.

Finally, the chiral quark model parameters are summarized in Table~\ref{model}. They have been fixed along the last two decades by thorough studies of hadron phenomenology such as meson~\cite{Segovia:2008zza, Segovia:2015dia, Ortega:2020uvc} and baryon~\cite{Valcarce:1995dm, Yang:2017qan, Yang:2019lsg} spectra, hadron decays and reactions~\cite{Segovia:2009zz, Segovia:2011zza, Segovia:2011dg}, coupling between conventional hadrons and hadron-hadron thresholds~\cite{Ortega:2009hj, Ortega:2016pgg, Ortega:2016hde} as well as molecular hadron-hadron formation~\cite{Ortega:2018cnm, Ortega:2021xst, Ortega:2022efc}.

\begin{table}[!t]
\caption{\label{model} Quark model parameters.}
\begin{ruledtabular}
\begin{tabular}{lrr}
Quark masses     & $m_u=m_d$ (MeV) &  313 \\
                 & $m_s$ (MeV)     & 555 \\
                 & $m_c$ (MeV)     & 1752 \\[2ex]
Goldstone bosons & $\Lambda_\pi=\Lambda_\sigma~$ (fm$^{-1}$) &   4.20 \\
                 & $\Lambda_\eta$ (fm$^{-1}$)     &   5.20 \\
                 & $g^2_{ch}/(4\pi)$                         &   0.54 \\
                 & $\theta_P(^\circ)$                        & -15 \\[2ex]
Confinement      & $a_c$ (MeV)         & 430\\
                 & $\mu_c$ (fm$^{-1})$ &   0.70\\
                 & $\Delta$ (MeV)      & 181.10 \\[2ex]
                 & $\alpha_0$              & 2.118 \\
                 & $\Lambda_0~$(fm$^{-1}$) & 0.113 \\
OGE              & $\mu_0~$(MeV)        & 36.976\\
                 & $\hat{r}_0~$(MeV~fm) & 28.170\\
\end{tabular}
\end{ruledtabular}
\end{table}

\begin{figure}[ht]
\epsfxsize=1.5in \epsfbox{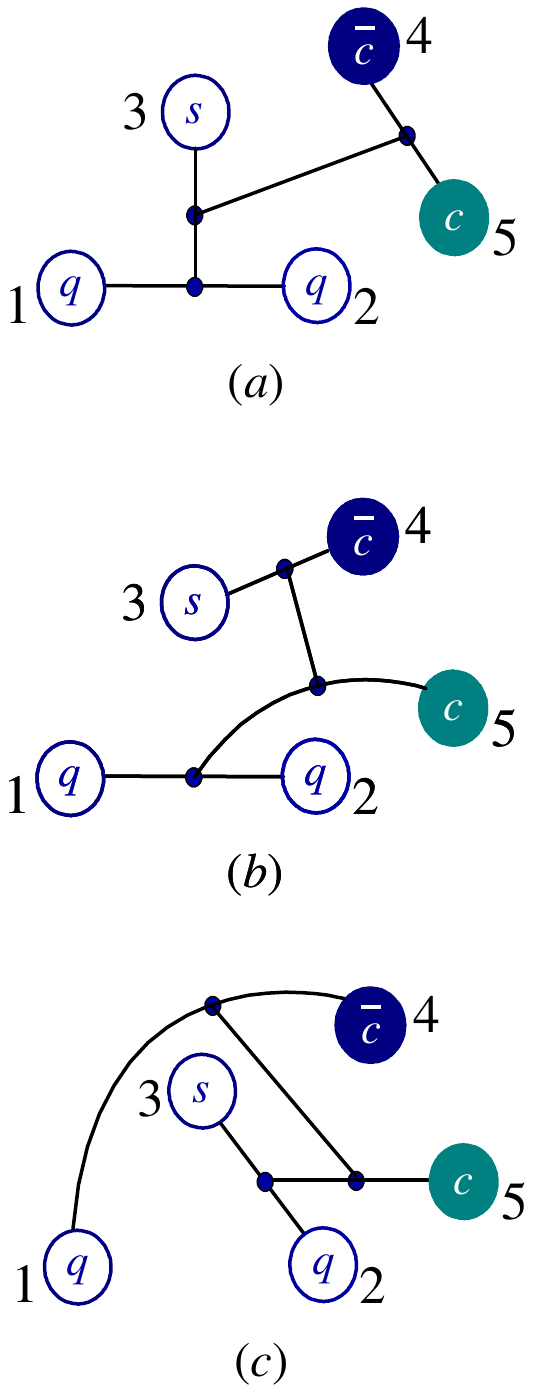}
\caption{Configurations of $qqsc\bar{c}$ $(q=u,\,d)$ pentaquarks.} \label{Pcs}
\end{figure}

\subsection{The wave function}

Three sets of configurations are generally needed for $qqsc\bar{c}~(q=u,\,d)$ pentaquarks and are shown in Fig.~\ref{Pcs}. Moreover, the anti-symmetry operator must be applied to each diagram as corresponding. Particularly, the anti-symmetry operator for the first configuration, panel (a) of Fig.~\ref{Pcs}, is
\begin{equation}
{\cal{A}}_1 = 1-(13)-(23) \,. \label{EE1}
\end{equation}
This expression also holds for the second case, panel (b) of Fig.~\ref{Pcs} $(b)$, \emph{viz} ${\cal{A}}_2={\cal{A}}_1$. Meanwhile, the anti-symmetry operator of panel (c) in Fig.~\ref{Pcs}, where the two heavy quarks are arranged in each sub-cluster, reads
\begin{equation}
{\cal{A}}_3 = 1-(12)-(13) \,. \label{EE2}
\end{equation}

The pentaquark wave function is a product of four terms: color, flavor, spin and space wave functions. Firstly, concerning the color degree-of-freedom, there are richer structure of multiquark systems than conventional hadrons. For instance, the color wave function of pentaquark must be colorless, but the way of reaching this condition can be done by either color-singlet, hidden-color or both at the same time. The authors of  Refs.~\cite{Harvey:1980rva, Vijande:2009kj} assert that it is enough to consider the color-singlet channel when all possible excited states of a system are included. However, a more economical and practical way is considering both, the color singlet wave function:
\begin{align}
\label{Color1}
\chi^{n c}_1 &= \frac{1}{\sqrt{18}}(rgb-rbg+gbr-grb+brg-bgr) \times \nonumber \\
&
\times (\bar r r+\bar gg+\bar bb) \,,
\end{align}
where $n=1,\,2,\,3$ is a label for each three different configurations in Fig.~\ref{Pcs} (it is of the same meaning for spin, flavor and space wave functions). They are in a common form but with different quark sequence, namely 123;45, 125;43 and 235;41, respectively. In matrix element calculation, one should switch the rest two cases into the first one of ascending order, and then the hidden-color wave function:
\begin{align}
\label{Color2}
\chi^{n c}_k &= \frac{1}{\sqrt{8}}(\chi^{n k}_{3,1}\chi_{2,8}-\chi^{n k}_{3,2}\chi_{2,7}-\chi^{n k}_{3,3}\chi_{2,6}+\chi^{n k}_{3,4}\chi_{2,5} \nonumber \\
& +\chi^{n k}_{3,5}\chi_{2,4}-\chi^{n k}_{3,6}\chi_{2,3}-\chi^{n k}_{3,7}\chi_{2,2}+\chi^{n k}_{3,8}\chi_{2,1}) \,,
\end{align}
where $k=2,\,3$ is an index which stands for the symmetric and anti-symmetric configuration of two identical quarks in the $3$-quark sub-cluster.

According to the SU(3) flavor symmetry in the isospin space, flavor wave functions for the sub-clusters mentioned above are given by:
\begin{align}
B^{1+}_{1, 0}  &= \frac{1}{\sqrt{12}}(2uds+2dus-usd-dsu-sud-sdu) \,, \\
B^{1-}_{1, 0}  &= \frac{1}{\sqrt{2}}(usd-sud+dsu-sdu) \,, \\
B^{1*}_{1, 0}  &= \frac{1}{\sqrt{6}}(uds+dus+usd+dsu+sud+sdu) \,, \\
B^{1+}_{0, 0} &= \frac{1}{\sqrt{2}}(usd-dsu+sud-sdu) \,, \\
B^{1-}_{0, 0} &= \frac{1}{\sqrt{12}}(2uds-2dus+usd-dsu-sud+sdu) \,,
\end{align}
\begin{align}
B^{2+}_{1, 0}  &= \frac{1}{\sqrt{2}}(ud+du)c \,, \\
B^{2-}_{0, 0} &= \frac{1}{\sqrt{2}}(ud-du)c \,, \\
B^{3\pm}_{\frac12, \frac12}  &= \frac{1}{\sqrt{2}}(us\pm su)c \,,  \,\, B^{3\pm}_{\frac12, -\frac12}  = \frac{1}{\sqrt{2}}(ds\pm sd)c  \,, \\
M^1_{0, 0} &= \bar{c}c \,,  \,\,\,\,\,  M^2_{0, 0} = \bar{c}s \,, \\
M^3_{\frac12, \frac12} &= \bar{c}u \,,  \,\,\,\,\,  M^3_{\frac12, -\frac12} = \bar{c}d \,,
\end{align}
where the superscript of cluster wave functions, $B$ and $M$, stand for the assigned number of each three configurations; and the subscripts refers to isospin ($I$) and its third component ($M_I$). Moreover, symbols of $+(*)$ and $-$ in the wave function stand for a symmetry and anti-symmetry property of two quarks in the $3$-quark sub-cluster. Consequently, flavor wave-functions for the 5-quark system with isospin $I=0$ and $1$ are
\begin{align}
&
\chi_{0, 0}^{1f1}(5) = B^{1 +}_{0, 0} M^1_{0, 0} \,, \,\,\,\,
\chi_{0, 0}^{1f2}(5) = B^{1 -}_{0, 0} M^1_{0, 0} \,,\\
&
\chi_{0, 0}^{2f1}(5) = B^{2 -}_{0, 0} M^2_{0, 0} \,,\\
&
\chi_{0, 0}^{3f1}(5) = \frac{1}{\sqrt{2}} (B^{3 +}_{\frac12, \frac12} M^3_{\frac12, -\frac12}-B^{3 +}_{\frac12, -\frac12} M^3_{\frac12, \frac12}) \,, \\
&
\chi_{0, 0}^{3f2}(5) = \frac{1}{\sqrt{2}} (B^{3 -}_{\frac12, \frac12} M^3_{\frac12, -\frac12}-B^{3 -}_{\frac12, -\frac12} M^3_{\frac12, \frac12}) \,, \\
&
\chi_{1, 0}^{1f1}(5) = B^{1 +}_{1, 0} M^1_{0, 0} \,, \,\,\,\,
\chi_{1, 0}^{1f2}(5) = B^{1 -}_{1, 0} M^1_{0, 0} \\
&
\chi_{1, 0}^{1f3}(5) = B^{1 *}_{1, 0} M^1_{0, 0} \,,  \,\,\,\,
\chi_{1, 0}^{2f1}(5) = B^{2 +}_{1, 0} M^2_{0, 0} \\
&
\chi_{1, 0}^{3f1}(5) = \frac{1}{\sqrt{2}} (B^{3 +}_{\frac12, \frac12} M^3_{\frac12, -\frac12}+B^{3 +}_{\frac12, -\frac12} M^3_{\frac12, \frac12}) \,, \\
&
\chi_{1, 0}^{3f2}(5) = \frac{1}{\sqrt{2}} (B^{3 -}_{\frac12, \frac12} M^3_{\frac12, -\frac12}+B^{3 -}_{\frac12, -\frac12} M^3_{\frac12, \frac12}) \,,
\end{align}
where the third component of isospin is set to be zero without loss of generality because there is no interaction in the Hamiltonian that can distinguish such a component.

We are going to consider herein a 5-quark system with total spin ranging from $1/2$ to $5/2$. The Hamiltonian does not have any spin-orbit coupling dependent interaction; therefore, the third component of spin is assumed to be equal to the total one without loss of generality. Then, the spin wave function is given by
\begin{align}
\label{Spin}
\chi_{\frac12,\frac12}^{n \sigma 1}(5) &= \chi_{\frac12,\frac12}^{n \sigma 1}(3) \chi_{00}^{\sigma} \,, \\
\chi_{\frac12,\frac12}^{n \sigma 2}(5) &= \chi_{\frac12,\frac12}^{n \sigma 2}(3) \chi_{00}^{\sigma} \,, \\
\chi_{\frac12,\frac12}^{n \sigma 3}(5) &= \sqrt{\frac{1}{3}} \chi_{\frac12,\frac12}^{n \sigma 1}(3) \chi_{10}^{\sigma} -\sqrt{\frac{2}{3}} \chi_{\frac12,-\frac12}^{n \sigma 1}(3) \chi_{11}^{\sigma} \,, \\
\chi_{\frac12,\frac12}^{n \sigma 4}(5) &= \sqrt{\frac{1}{3}} \chi_{\frac12,\frac12}^{n \sigma 2}(3) \chi_{10}^{\sigma} - \sqrt{\frac{2}{3}} \chi_{\frac12,-\frac12}^{n \sigma 2}(3) \chi_{11}^{\sigma} \,, \\
\chi_{\frac12,\frac12}^{n \sigma 5}(5) &= \sqrt{\frac{1}{6}} \chi_{\frac32,-\frac12}^{n \sigma}(3) \chi_{11}^{\sigma}
-\sqrt{\frac{1}{3}} \chi_{\frac32,\frac12}^{n \sigma}(3) \chi_{10}^{\sigma} \nonumber \\
&
+\sqrt{\frac{1}{2}} \chi_{\frac32,\frac32}^{n \sigma}(3) \chi_{1-1}^{\sigma} \,,
\end{align}
for $S=1/2$, and
\begin{align}
\chi_{\frac32,\frac32}^{n \sigma 1}(5) &= \chi_{\frac12,\frac12}^{n \sigma 1}(3) \chi_{11}^{\sigma} \,, \\
\chi_{\frac32,\frac32}^{n \sigma 2}(5) &= \chi_{\frac12,\frac12}^{n \sigma 2}(3) \chi_{11}^{\sigma} \,, \\
\chi_{\frac32,\frac32}^{n \sigma 3}(5) &= \chi_{\frac32,\frac32}^{n \sigma}(3) \chi_{00}^{\sigma} \,, \\
\chi_{\frac32,\frac32}^{n \sigma 4}(5) &= \sqrt{\frac{3}{5}}
\chi_{\frac32,\frac32}^{n \sigma}(3) \chi_{10}^{\sigma} -\sqrt{\frac{2}{5}} \chi_{\frac32,\frac12}^{n \sigma}(3) \chi_{11}^{\sigma} \,, \\
\end{align}
for $S=3/2$, and
\begin{align}
\chi_{\frac52,\frac52}^{n \sigma 1}(5) &= \chi_{\frac32,\frac32}^{n \sigma}(3) \chi_{11}^{\sigma} \,,
\end{align}
for $S=5/2$. These expressions can be obtained easily through considering the 3-quark and quark-antiquark sub-clusters, and using a SU(2) algebra.

Among the different methods for solving a complex Schr\"odinger-like 5-body bound state equation, we use the Rayleigh-Ritz variational principle, which is one of the most extended tool to solve eigenvalue problems due to its simplicity and flexibility. Then, the spatial wave function of a $5$-quark system is written as follows:
\begin{align}
\label{eq:WFexp}
&
\psi_{LM_L}=[ [ [ \phi_{n_1l_1}(\vec{\rho} e^{i\theta})\phi_{n_2l_2}(\vec{\lambda} e^{i\theta})]_{l} \phi_{n_3l_3}(\vec{r} e^{i\theta}) ]_{l^{\prime}} \nonumber\\
&
\hspace*{1.60cm}  \phi_{n_4l_4}(\vec{R} e^{i\theta}) ]_{LM_L} \,.
\end{align}
Taken the first configuration of Fig.~\ref{Pcs} $(a)$ as an example, the internal Jacobi coordinates are defined as
\begin{align}
\vec{\rho} &= \vec{x}_1-\vec{x}_2 \,, \\
\vec{\lambda} &= \vec{x}_3 - (\frac{{m_1\vec{x}}_1+{m_2\vec{x}}_2}{m_1+m_2}) \,,  \\
\vec{r} &= \vec{x}_4-\vec{x}_5 \,, \\
\vec{R} &= \left(\frac{{m_1\vec{x}}_1+{m_2\vec{x}}_2 + {m_3\vec{x}}_3}{m_1+m_2+m_3}\right) \nonumber \\
&
- \left(\frac{{m_4\vec{x}}_4+{m_5\vec{x}}_5}{m_4+m_5}\right) \,.
\end{align}
The other two configurations of Fig.~\ref{Pcs}, \emph{i.e.} panels $(b)$ and $(c)$, are very similar but considering a different arrangement of quark sequence. This choice is convenient because the center-of-mass kinetic term $T_{CM}$ can be completely eliminated for a nonrelativistic system and it also allows us to extend the coordinates of relative motion between quarks into the complex scaling ground.

It is important how to choose the basis on which to expand the genuine wave function of Eq.~\eqref{eq:WFexp}. Herein, by employing the Gaussian expansion method (GEM)~\cite{Hiyama:2003cu}, spatial wave functions of each four relative motions are all expanded with Gaussian basis functions, whose sizes are taken in geometric progressions. This method has proven to be quite efficient on solving the bound-state problem of a multiquark systems~\cite{Yang:2015bmv, Yang:2020twg, Yang:2019itm}, and details on how the geometric progression is fixed can be found in \emph{e.g} Ref.~\cite{Yang:2015bmv}. Accordingly, the form of orbital wave functions $\phi$ in Eq.~\eqref{eq:WFexp} reads
\begin{align}
&
\phi_{nlm}(\vec{r} e^{i\theta}\,)=N_{nl} (r e^{i\theta})^{l} e^{-\nu_{n} (r e^{i\theta})^2} Y_{lm}(\hat{r})\,.
\end{align}
Since only $S$-wave states of $qqsc\bar{c}$ pentaquarks are investigated in this work, the spherical harmonic function is just a constant, \emph{viz.} $Y_{00}=\sqrt{1/4\pi}$, and thus no laborious Racah algebra is needed while computing matrix elements.

Finally, in order to fulfill the Pauli principle, the complete anti-symmetric complex wave-function can be written as
\begin{align}
\label{TPs}
 \Psi_{J M_J, I} &= \sum_{i, j, k} c_{ijk} \Psi_{J M_J, I, i, j, k} \nonumber \\
 &=\sum_{i, j, k} \sum_{n=1}^3 c_{ijk} {\cal A}_n \left[ \left[ \psi_{L M_L} \chi^{n \sigma_i}_{S M_S}(5) \right]_{J M_J} \chi^{n f_j}_I \chi^{n c}_k \right] \,,
\end{align}
where ${\cal A}_n$ is the antisymmetry operator of a 5-quark system and their expressions are shown in Eq.~(\ref{EE1}) and Eq.~(\ref{EE2}), respectively. This is needed because we have constructed an antisymmetric wave function for only two light quarks in the baryon sub-cluster, the remaining quark of system has been added to the wave function by simply considering appropriate Clebsch-Gordan coefficients. Furthermore, the so-called expansion coefficient, $c_{ijk}$, fulfills
\begin{align}
\label{TPs}
  \vert c_{ijk} \vert^2&=\langle \Psi_{J M_J, I, i, j, k} \vert \Psi_{JM_J,I} \rangle \,,  \\
  \sum_{i,j,k} \vert c_{ijk} \vert^2&=1 \,.
\end{align}
They are determined, together with the pentaquark eigenenergy, by a generalized matrix eigenvalue problem.

In the next section, where our results on hidden-charm pentaquarks with strangeness are discussed, we firstly study the systems by a real-range analysis, \emph{viz.}, the rotated angle $\theta$ is equal to $0^{\circ}$. In this case, when a complete coupled-channel calculation of matrix diagonalization is performed, possible resonant states are embedded in the continuum. However, one can employ the CSM, with appropriate non-zero values of $\theta$, to disentangle bound, resonance and scattering states in a complex energy plane. Accordingly, with the purpose of solving manageable eigevalue problems, the artificial parameter of rotated angle is ranged form $0^\circ$ to $6^\circ$. Meanwhile, with the cooperation of real- and complex-range computations, available exotic states, which are firstly obtained within a complex-range analysis, and then can be identified among continuum states according to its mass in a real-range calculation, are further investigated by analyzing their dominant quark arrangements, sizes and decay patterns.


\section{RESULTS}
\label{sec:results}

\begin{table*}[!t]
\caption{\label{GCC1} All possible channels for $qqsc\bar{c}$ pentaquark systems with $J^P=1/2^-$. Each channel is assigned an index in the second column, it reflects a particular combination of spin ($\chi_J^{n \sigma_i}$), flavor ($\chi_I^{n f_j}$) and color ($\chi_k^{n c}$) wave functions that are shown explicitly in the third and fifth column. Baryon-meson configuration is listed in the fourth and last column, the superscripts 1 and 8 stand for color-singlet and -octet state, respectively.}
\begin{ruledtabular}
\begin{tabular}{cccccc}
& & \multicolumn{2}{c}{$I=0$} & \multicolumn{2}{c}{$I=1$} \\
$J^P$~~&~Index~ & $\chi_J^{n \sigma_i}$;~$\chi_I^{n f_j}$;~$\chi_k^{n c}$; & Channel~~ & $\chi_J^{n \sigma_i}$;~$\chi_I^{n f_j}$;~$\chi_k^{n c}$; & Channel~~ \\
&&$[i; ~j; ~k;~n]$& &$[i; ~j; ~k;~n]$&  \\[2ex]
$\frac{1}{2}^-$ & 1  & $[1, 2;~1, 2;~1;~1]$   & $(\Lambda \eta_c)^1$ & $[1,2; ~1,2; ~1; ~1]$ & $(\Sigma \eta_c)^1$\\
& 2  & $[2;~1;~1;~2]$  & $(\Lambda_c D_s)^1$ & $[3,4; ~1,2; ~1; ~1]$  & $(\Sigma J/\psi)^1$ \\
& 3  & $[1;~1;~1;~3]$     & $(\Xi'_c D)^1$  & $[5; ~3; ~1; ~1]$   & $(\Sigma^* J/\psi)^1$ \\
& 4  & $[2;~2;~1;~3]$  & $(\Xi_c D)^1$ & $[1; ~1; ~1; ~2]$ & $(\Sigma_c D_s)^1$ \\
& 5  & $[3,4;~1, 2;~1;~1]$     & $(\Lambda J/\psi)^1$  & $[3; ~1; ~1; ~2]$   & $(\Sigma_c D^*_s)^1$ \\
& 6  & $[4;~1;~1;~2]$  & $(\Lambda_c D^*_s)^1$  & $[5; ~1; ~1; ~2]$   & $(\Sigma^*_c D^*_s)^1$  \\
& 7  & $[3;~1;~1;~3]$     & $(\Xi'_c D^*)^1$  & $[1; ~1; ~1; ~3]$   & $(\Xi'_c D)^1$ \\
& 8  & $[4;~2;~1;~3]$  & $(\Xi_c D^*)^1$  & $[2; ~2; ~1; ~3]$   & $(\Xi_c D)^1$  \\
& 9  & $[5;~1;~1;~3]$   & $(\Xi^*_c D^*)^1$   & $[3; ~1; ~1; ~3]$  & $(\Xi'_c D^*)^1$ \\
& 10 & $[1,2;~1, 2;~2, 3;~1]$  & $(\Lambda \eta_c)^8$ & $[4; ~2; ~1; ~3]$   &  $(\Xi_c D^*)^1$  \\
& 11  & $[1,2;~1;~2, 3;~2]$   & $(\Lambda_c D_s)^8$   & $[5; ~1; ~1; ~3]$  & $(\Xi^*_c D^*)^1$ \\
& 12 & $[1,2;~1;~2, 3;~3]$  & $(\Xi'_c D)^8$  & $[1,2;~1, 2;~2, 3;~1]$  & $(\Sigma \eta_c)^8$ \\
& 13  & $[1,2;~2;~2, 3;~3]$   & $(\Xi_c D)^8$ & $[3,4;~1, 2;~2, 3;~1]$  & $(\Sigma J/\psi)^8$ \\
& 14 & $[3,4;~1, 2;~2, 3;~1]$  & $(\Lambda J/\psi)^8$  & $[5;~1, 2;~2, 3;~1]$  & $(\Sigma^* J/\psi)^8$ \\
& 15  & $[3,4;~1;~2, 3;~2]$   & $(\Lambda_c D^*_s)^8 $ & $[1,2;~1;~2, 3;~2]$  & $(\Sigma_c D_s)^8$  \\
& 16 & $[3,4;~1;~2, 3;~3]$  & $(\Xi'_c D^*)^8$  & $[3,4;~1;~2, 3;~2]$  & $(\Sigma_c D^*_s)^8$  \\
& 17  & $[3,4;~2;~2, 3;~3]$   & $(\Xi_c D^*)^8$ & $[5;~1;~3;~2]$  & $(\Sigma^*_c D^*_s)^8$  \\
& 18 & $[5;~1, 2;~2, 3;~3]$  & $(\Xi^*_c D^*)^8$  & $[1,2;~1;~2, 3;~3]$  & $(\Xi'_c D)^8$  \\
& 19 & & & $[1,2;~2;~2, 3;~3]$ & $(\Xi_c D)^8$ \\
& 20 & & & $[3,4;~1;~2, 3;~3]$ & $(\Xi'_c D^*)^8$ \\
& 21 & & & $[3,4;~2;~2, 3;~3]$ & $(\Xi_c D^*)^8$ \\
& 22 & & & $[5;~1,2;~2, 3;~3]$ & $(\Xi^*_c D^*)^8$ \\
\end{tabular}
\end{ruledtabular}
\end{table*}

\begin{table*}[!t]
\caption{\label{GCC2} All possible channels for $qqsc\bar{c}$ pentaquark systems with $3/2^-$ and $5/2^-$. Columns are similarly organized as those in Table~\ref{GCC1}.}
\begin{ruledtabular}
\begin{tabular}{cccccc}
& & \multicolumn{2}{c}{$I=0$} & \multicolumn{2}{c}{$I=1$} \\
$J^P$~~&~Index~ & $\chi_J^{n \sigma_i}$;~$\chi_I^{n f_j}$;~$\chi_k^{n c}$; & Channel~~ & $\chi_J^{n \sigma_i}$;~$\chi_I^{n f_j}$;~$\chi_k^{n c}$; & Channel~~ \\
&&$[i; ~j; ~k;~n]$& &$[i; ~j; ~k;~n]$&  \\[2ex]
$\frac{3}{2}^-$ & 1 & $[1,2; ~1,2; ~1; ~1]$  & $(\Lambda J/\psi)^1$ & $[1,2; ~1,2; ~1; ~1]$ & $(\Sigma J/\psi)^1$ \\
& 2  & $[2; ~1; ~1; ~2]$  & $(\Lambda_c D^*_s)^1$ & $[3; ~3; ~1; ~1]$  & $(\Sigma^* \eta_c)^1$ \\
& 3  & $[1; ~1; ~1; ~3]$   & $(\Xi'_c D^*)^1$   & $[4; ~3; ~1; ~1]$   & $(\Sigma^* J/\psi)^1$ \\
& 4  & $[2; ~2; ~1; ~3]$   & $(\Xi_c D^*)^1$  & $[1; ~1; ~1; ~2]$ & $(\Sigma_c D^*_s)^1$ \\
& 5  & $[3; ~1; ~1; ~3]$   & $(\Xi^*_c D)^1$  & $[3; ~1; ~1; ~2]$  & $(\Sigma^*_c D_s)^1$  \\
& 6  & $[4; ~1; ~1; ~3]$   & $(\Xi^*_c D^*)^1$  & $[4; ~1; ~1; ~2]$  & $(\Sigma^*_c D^*_s)^1$ \\
& 7 & $[1,2; ~1,2; ~2,3; ~1]$  & $(\Lambda J/\psi)^8$ & $[1; ~1; ~1; ~3]$ & $(\Xi'_c D^*)^1$ \\
& 8  & $[1,2; ~1; ~2,3; ~2]$  & $(\Lambda_c D^*_s)^8$ & $[2; ~2; ~1; ~3]$  & $(\Xi_c D^*)^1$ \\
& 9  & $[1,2; ~1; ~2,3; ~3]$   & $(\Xi'_c D^*)^8$   & $[3; ~1; ~1; ~3]$   & $(\Xi^*_c D)^1$ \\
& 10  & $[1,2; ~2; ~2,3; ~3]$   & $(\Xi_c D^*)^8$  & $[4; ~1; ~1; ~3]$ & $(\Xi^*_c D^*)^1$ \\
& 11  & $[3; ~1,2; ~2,3; ~3]$   & $(\Xi^*_c D)^8$  & $[1,2; ~1,2; ~2,3; ~1]$  & $(\Sigma J/\psi)^8$ \\
& 12  & $[4; ~1,2; ~2,3; ~3]$   & $(\Xi^*_c D^*)^8$  & $[3; ~1,2; ~2,3; ~1]$  & $(\Sigma^* \eta_c)^8$  \\
& 13 & & & $[4; ~1,2; ~2,3; ~1]$ & $(\Sigma^* J/\psi)^8$ \\
& 14 & & & $[1,2; ~1; ~2,3; ~2]$ & $(\Sigma_c D^*_s)^8$ \\
& 15 & & & $[3; ~1; ~3; ~2]$ & $(\Sigma^*_c D_s)^8$ \\
& 16 & & & $[4; ~1; ~3; ~2]$ & $(\Sigma^*_c D^*_s)^8$ \\
& 17 & & & $[1,2; ~1; ~2,3; ~3]$ & $(\Xi'_c D^*)^8$ \\
& 18 & & & $[1,2; ~2; ~2,3; ~3]$ & $(\Xi_c D^*)^8$ \\
& 19 & & & $[3; ~1,2; ~2,3; ~3]$ & $(\Xi^*_c D)^8$ \\
& 20 & & & $[4; ~1,2; ~2,3; ~3]$ & $(\Xi^*_c D^*)^8$ \\ [2ex]
$\frac{5}{2}^-$ & 1 & $[1; ~1; ~1; ~3]$  & $(\Xi^*_c D^*)^1$ & $[1; ~3; ~1; ~1]$ & $(\Sigma^* J/\psi)^1$ \\
& 2 & $[1; ~1,2; ~2,3; ~3]$  & $(\Xi^*_c D^*)^8$ & $[1; ~1; ~1; ~2]$ & $(\Sigma^*_c D^*_s)^1$ \\
& 3 & & & $[1; ~1; ~1; ~3]$ & $(\Xi^*_c D^*)^1$ \\
& 4 & & & $[1; ~1,2; ~2,3; ~1]$ & $(\Sigma^* J/\psi)^8$ \\
& 5 & & & $[1; ~1; ~3; ~2]$ & $(\Sigma^*_c D^*_s)^8$ \\
& 6 & & & $[1; ~1,2; ~2,3; ~3]$ & $(\Xi^*_c D^*)^8$ \\
\end{tabular}
\end{ruledtabular}
\end{table*}

The lowest-lying and possible resonant states of $S$-wave $qqsc\bar{c}$ pentaquarks are investigated by taking into account three types of baryon-meson configurations, which includes the $(qqs)(\bar{c}c)$, $(qqc)(\bar{c}s)$ and $(qsc)(\bar{c}c)$, and they are shown in Fig.~\ref{Pcs}. Therein, the angular momenta $l_1$, $l_2$, $l_3$, $l_4$, which appear in Eq.~\eqref{eq:WFexp}, are all equal to zero. Therefore, the total angular momentum, $J$, coincides with the total spin, $S$, and can take values $1/2$, $3/2$ and $5/2$, respectively. The parity of pentaquark system is then negative. Table~\ref{GCC1} and~\ref{GCC2} list all allowed baryon-meson configurations of each $I(J^P)$-channel. In particular, channels are indexed in the second column, the third and fifth columns present the necessary basis combination in spin $(\chi^{n \sigma_i}_J)$, flavor $(\chi^{n f_j}_I)$, and color $(\chi^{n c}_k)$ degrees-of-freedom along with possible configurations $(n=1, 2, 3)$. Physical channels with color-singlet (labeled with the super-index $1$) and color-octet (labeled with the super-index $8$) configurations are listed in the fourth and sixth columns, respectively.

First of all, the lowest-lying $qqsc\bar{c}$ pentaquark in each channel is computed with a rotated angle $\theta=0^\circ$. Tables~\ref{Gresult1},~\ref{Gresult2},~\ref{Gresult3},~\ref{Gresult4},~\ref{Gresult5} and~\ref{Gresult6} summarize our calculated lowest masses of the $qqsc\bar{c}$ system with spin-parity $J^P=\frac{1}{2}^-$, $\frac{3}{2}^-$ and $\frac{5}{2}^-$, isospin $I=0$ and $1$, respectively. In these tables, baryon-meson configuration is listed in the first column, the superscripts 1 and 8 stand for color-singlet and -octet states, respectively. The experimental threshold value of a baryon-meson channel is then listed in the parenthesis. The lowest theoretical mass obtained in each channel is shown in the second column, and the binding energy is presented in the following one. A mixture of color-singlet and -octet configurations for each baryon-meson case is considered, and the coupled mass and binding energy is shown in the last column. The lowest-lying mass in coupled-channels calculation which includes all-color-singlet, all-color-octet and a complete coupled-channel one  indicated at the bottom of tables.

\begin{figure}[!t]
\includegraphics[clip, trim={3.0cm 1.9cm 3.0cm 1.0cm}, width=0.45\textwidth]{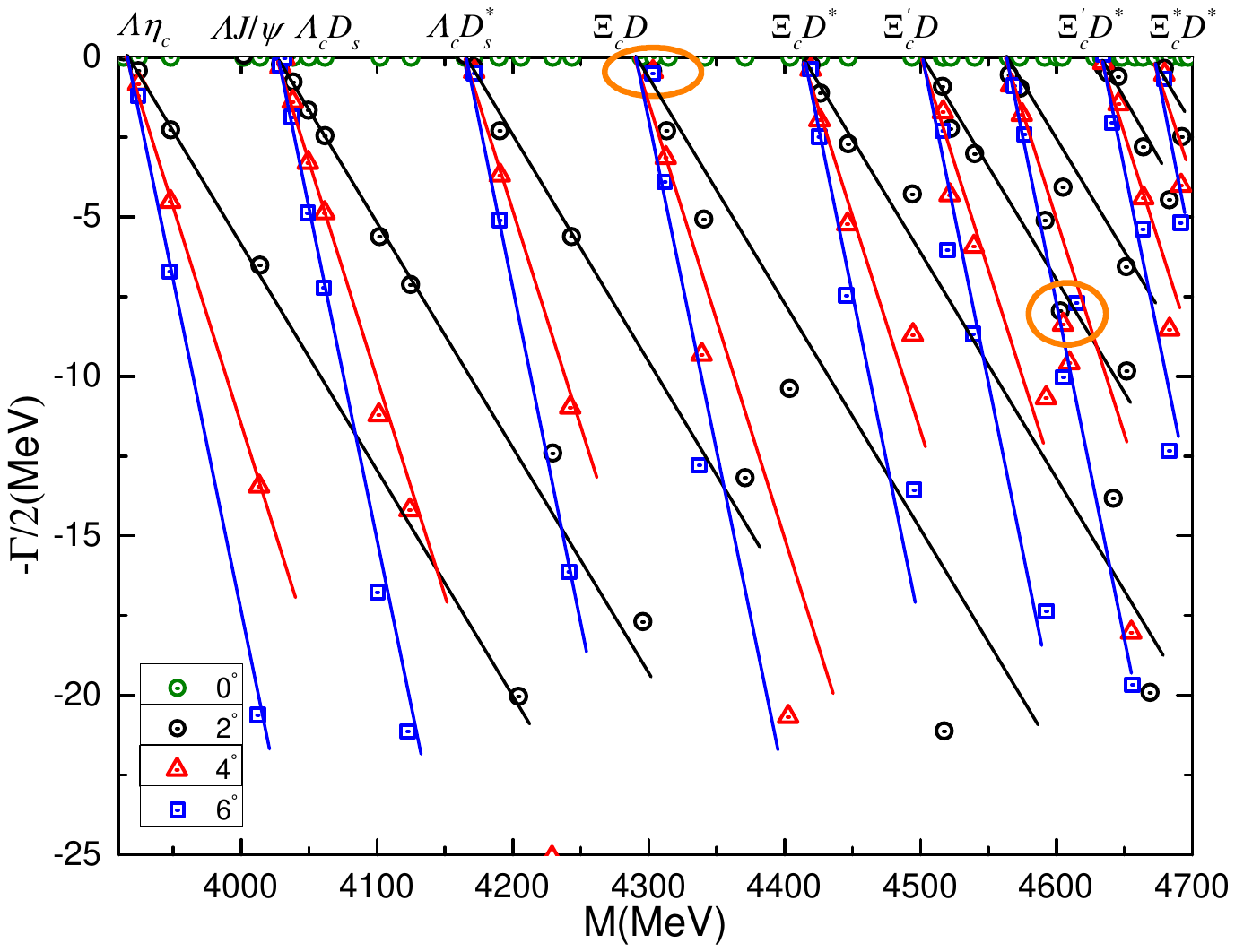}
\caption{ Complex energies of complete coupled-channel calculation for the $qqsc\bar{c}$ pentaquark within $IJ^P=0(\frac12^-)$.} \label{PP1}
\end{figure}

\begin{figure}[!t]
\includegraphics[clip, trim={3.0cm 1.9cm 3.0cm 1.0cm}, width=0.45\textwidth]{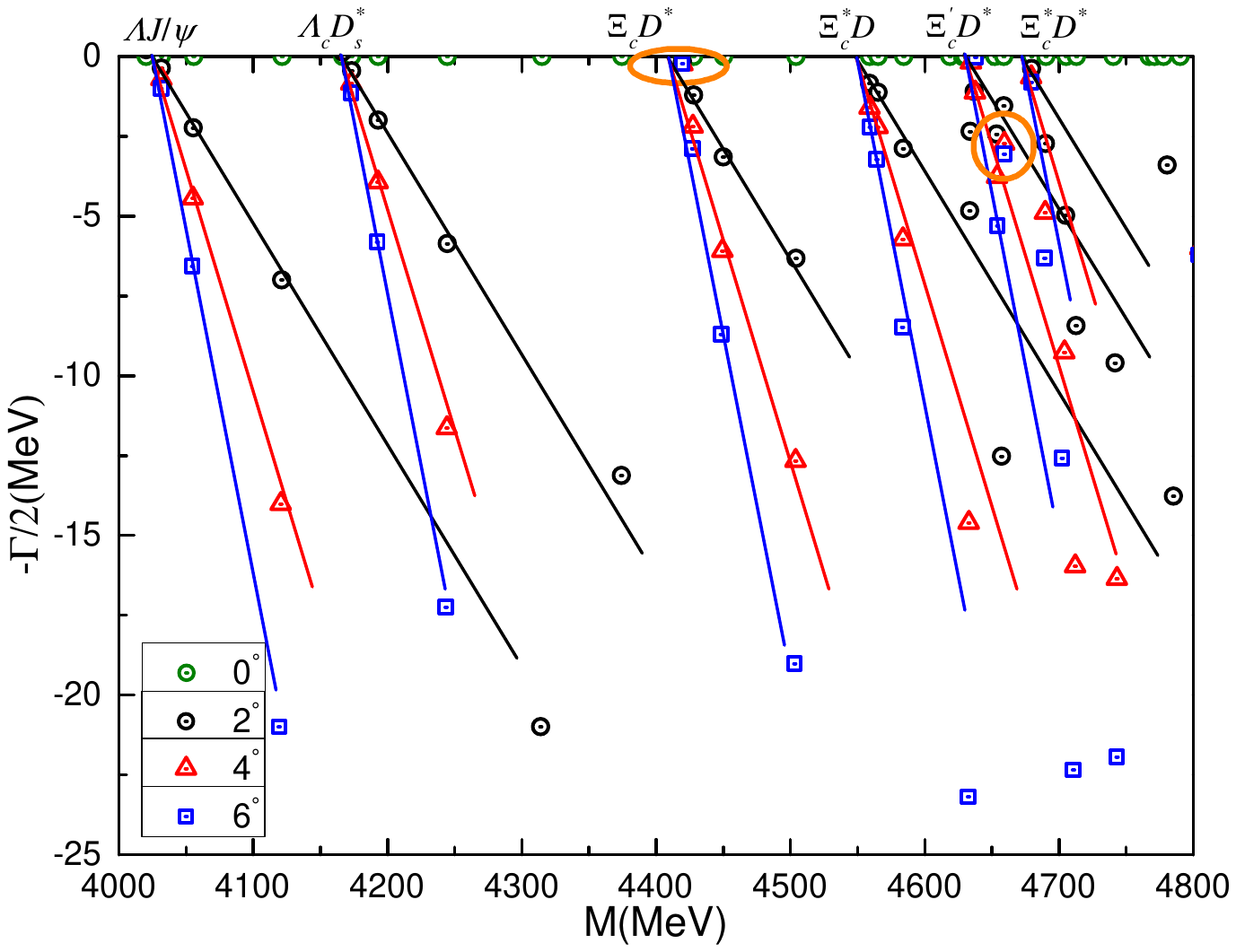}
\caption{ Complex energies of complete coupled-channel calculation for the $qqsc\bar{c}$ pentaquark within $IJ^P=0(\frac32^-)$.} \label{PP2}
\end{figure}

\begin{figure}[!t]
\includegraphics[clip, trim={3.0cm 1.9cm 3.0cm 1.0cm}, width=0.45\textwidth]{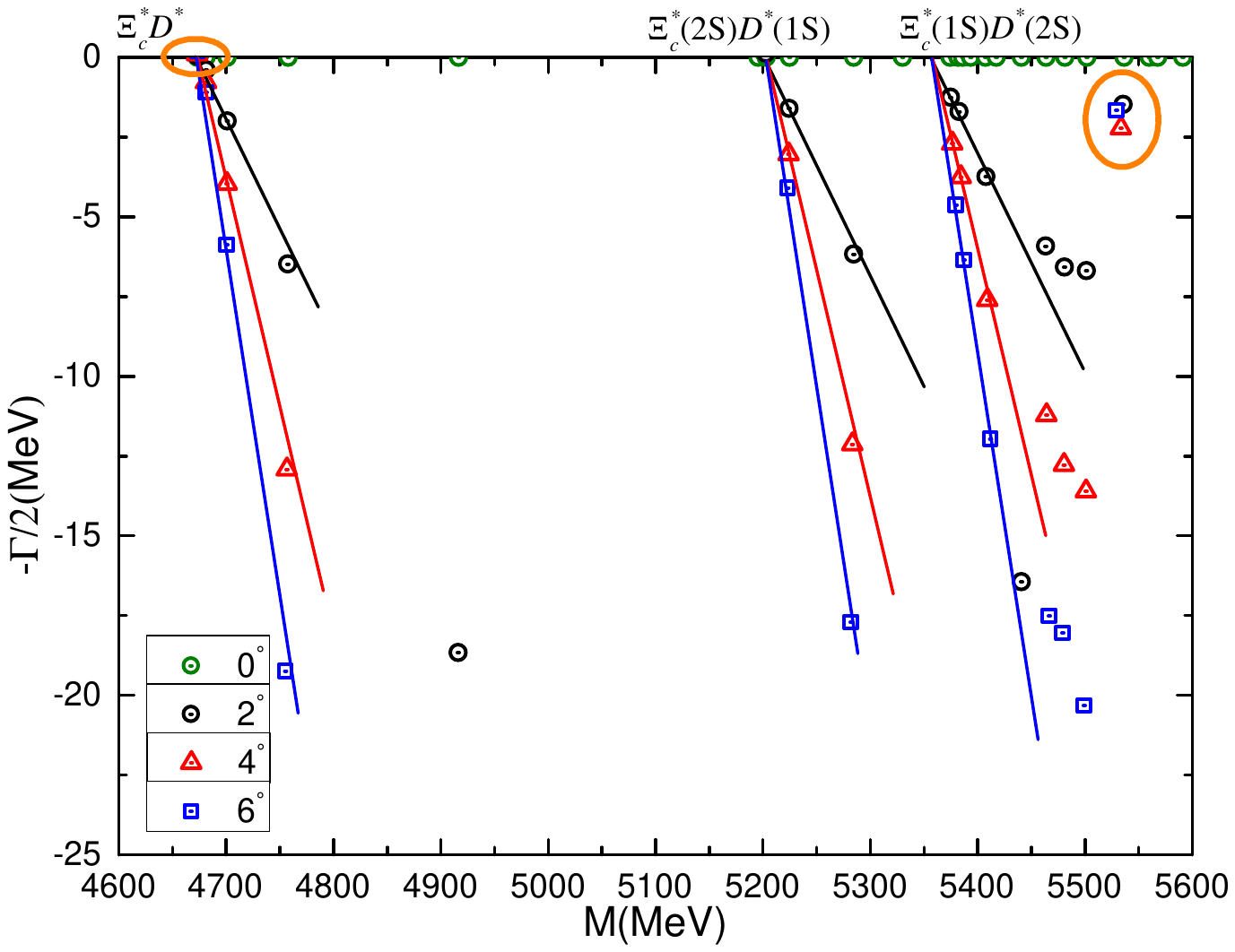}
\caption{ Complex energies of complete coupled-channel calculation for the $qqsc\bar{c}$ pentaquark within $IJ^P=0(\frac52^-)$.} \label{PP3}
\end{figure}

\begin{figure}[!t]
\includegraphics[clip, trim={3.0cm 1.9cm 3.0cm 1.0cm}, width=0.45\textwidth]{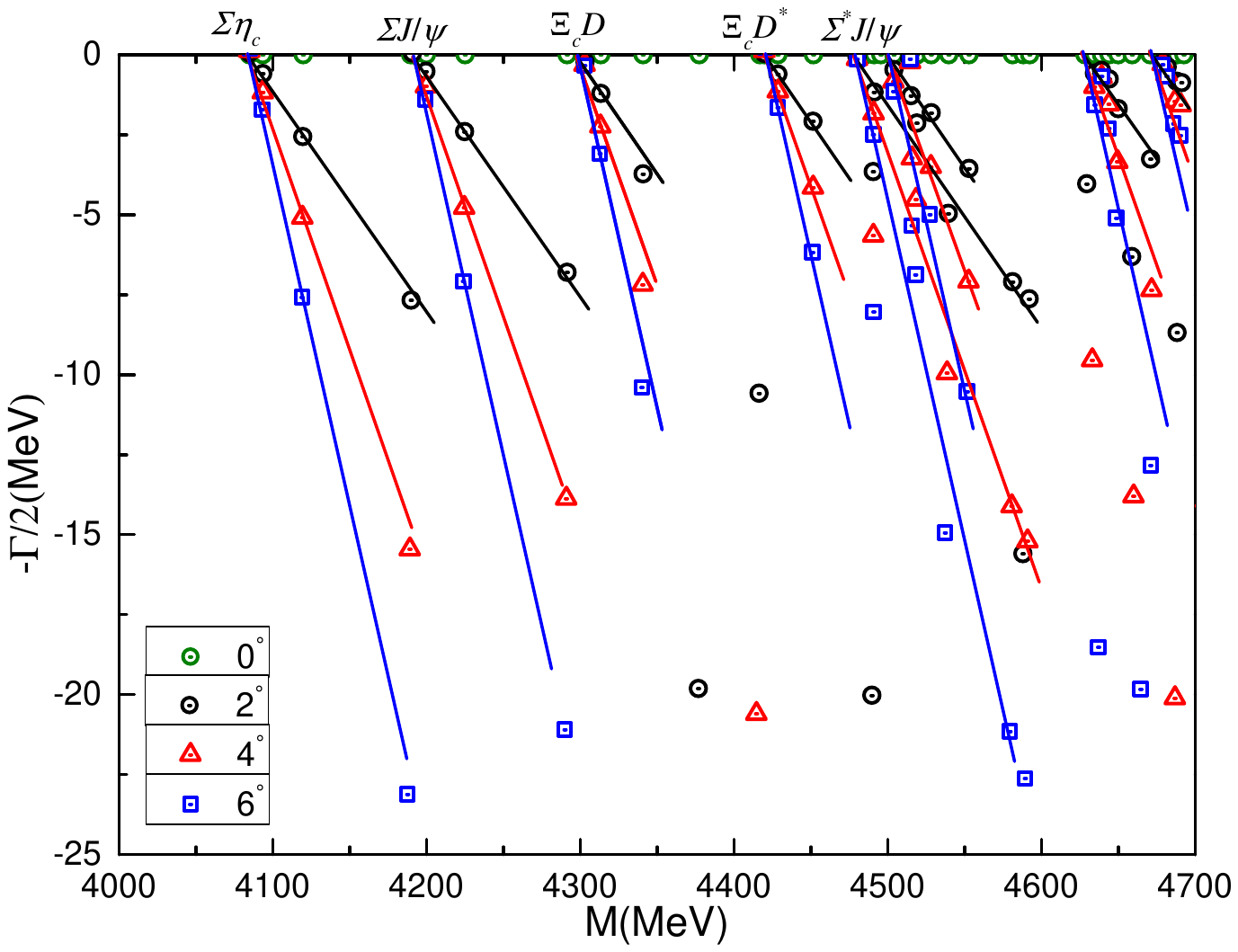}\\
\includegraphics[clip, trim={3.0cm 1.9cm 3.0cm 1.0cm}, width=0.45\textwidth]{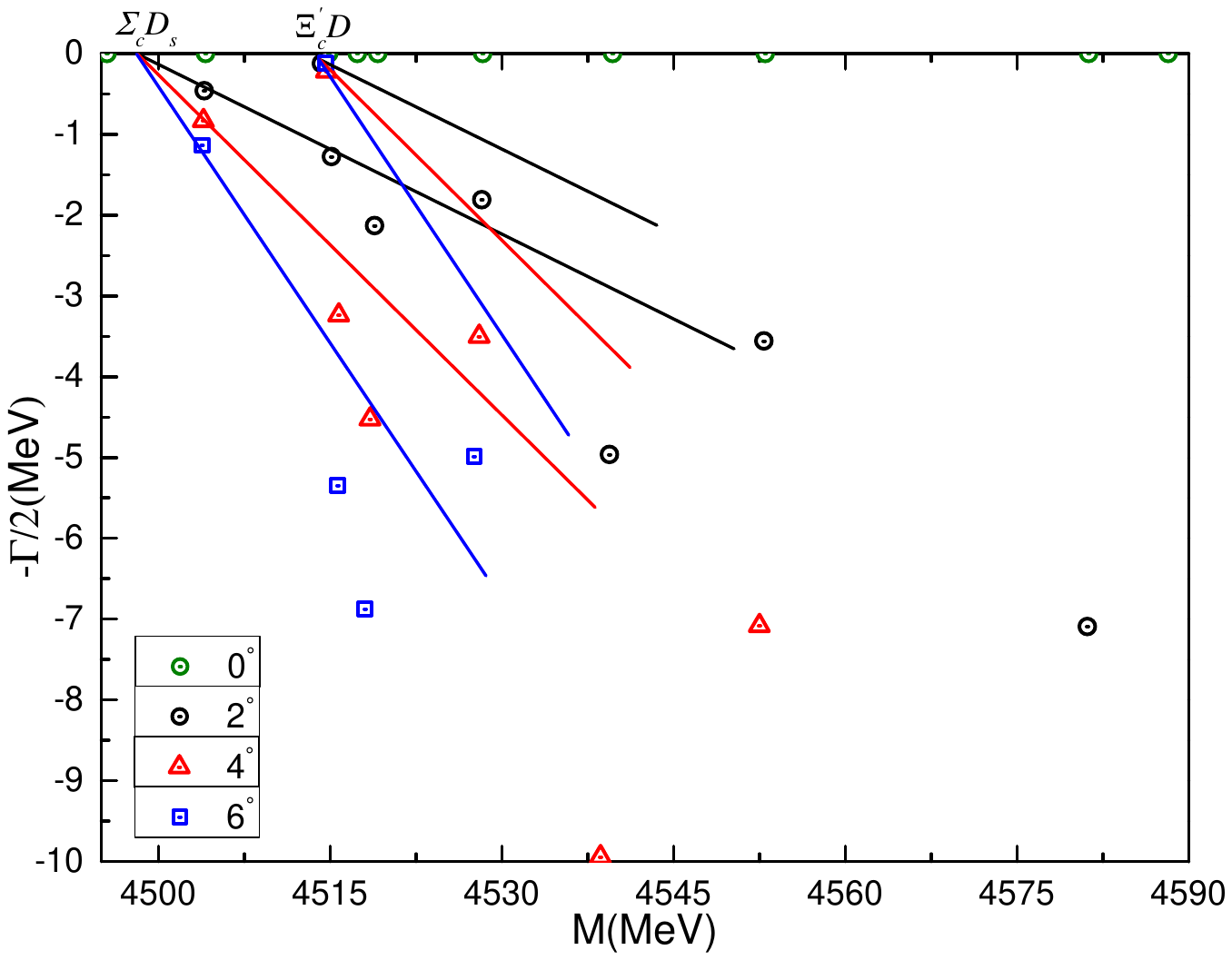}\\
\includegraphics[clip, trim={3.0cm 1.9cm 3.0cm 1.0cm}, width=0.45\textwidth]{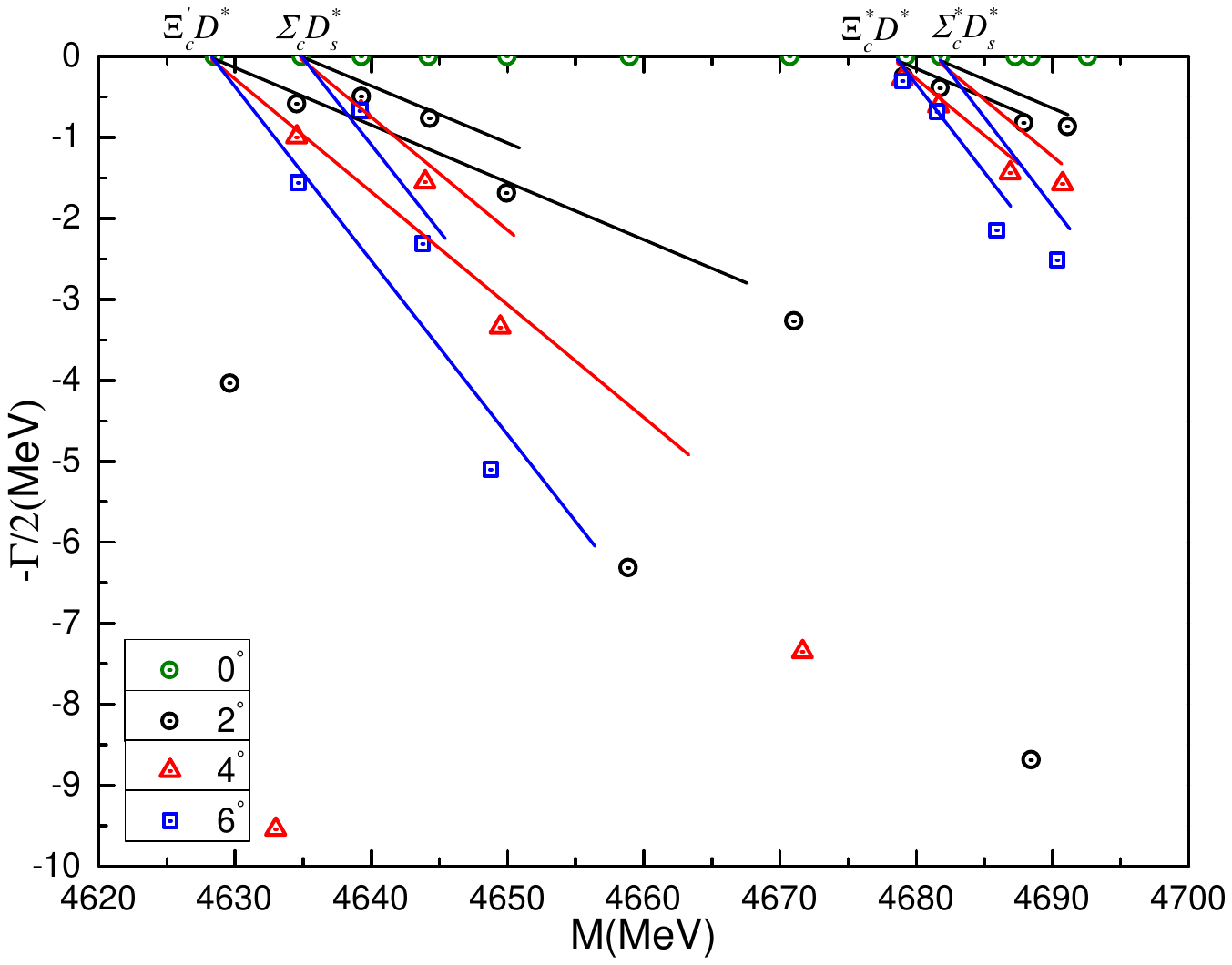}
\caption{ {\it Top panel:} Complex energies of complete coupled-channel calculation for the $qqsc\bar{c}$ pentaquark within $IJ^P=1(\frac12^-)$. {\it Middle panel:} Enlarged top panel, with real values of energy ranging from $4.49\,\text{GeV}$ to $4.59\,\text{GeV}$. {\it Bottom panel:} Enlarged top panel, with real values of energy ranging from $4.62\,\text{GeV}$ to $4.70\,\text{GeV}$.} \label{PP4}
\end{figure}

\begin{figure}[!t]
\includegraphics[clip, trim={3.0cm 1.9cm 3.0cm 0.75cm}, width=0.45\textwidth]{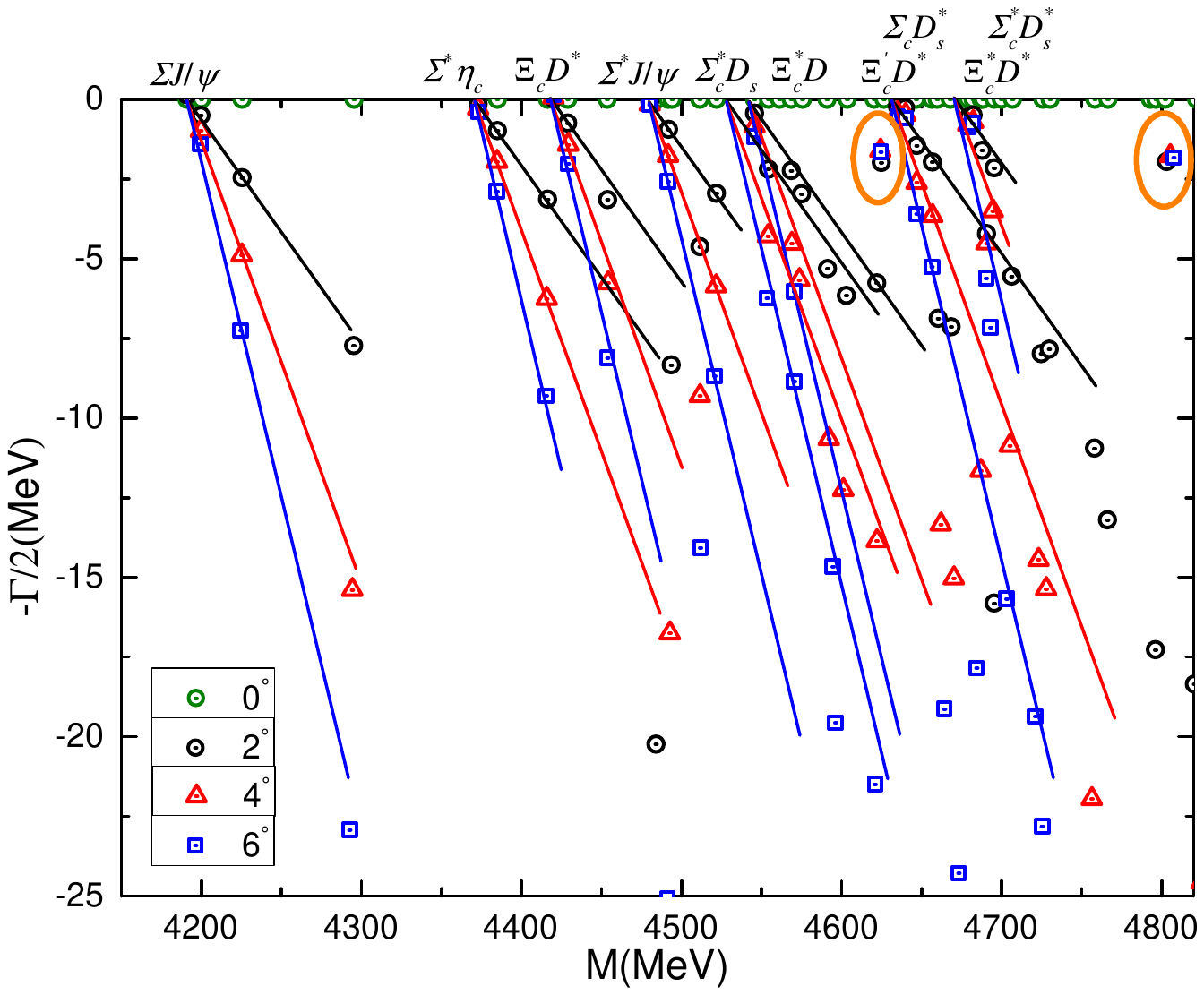}
\caption{ Complex energies of complete coupled-channel calculation for the $qqsc\bar{c}$ pentaquark within $IJ^P=1(\frac32^-)$.} \label{PP5}
\end{figure}

\begin{figure}[!t]
\includegraphics[clip, trim={3.0cm 1.9cm 3.0cm 1.0cm}, width=0.45\textwidth]{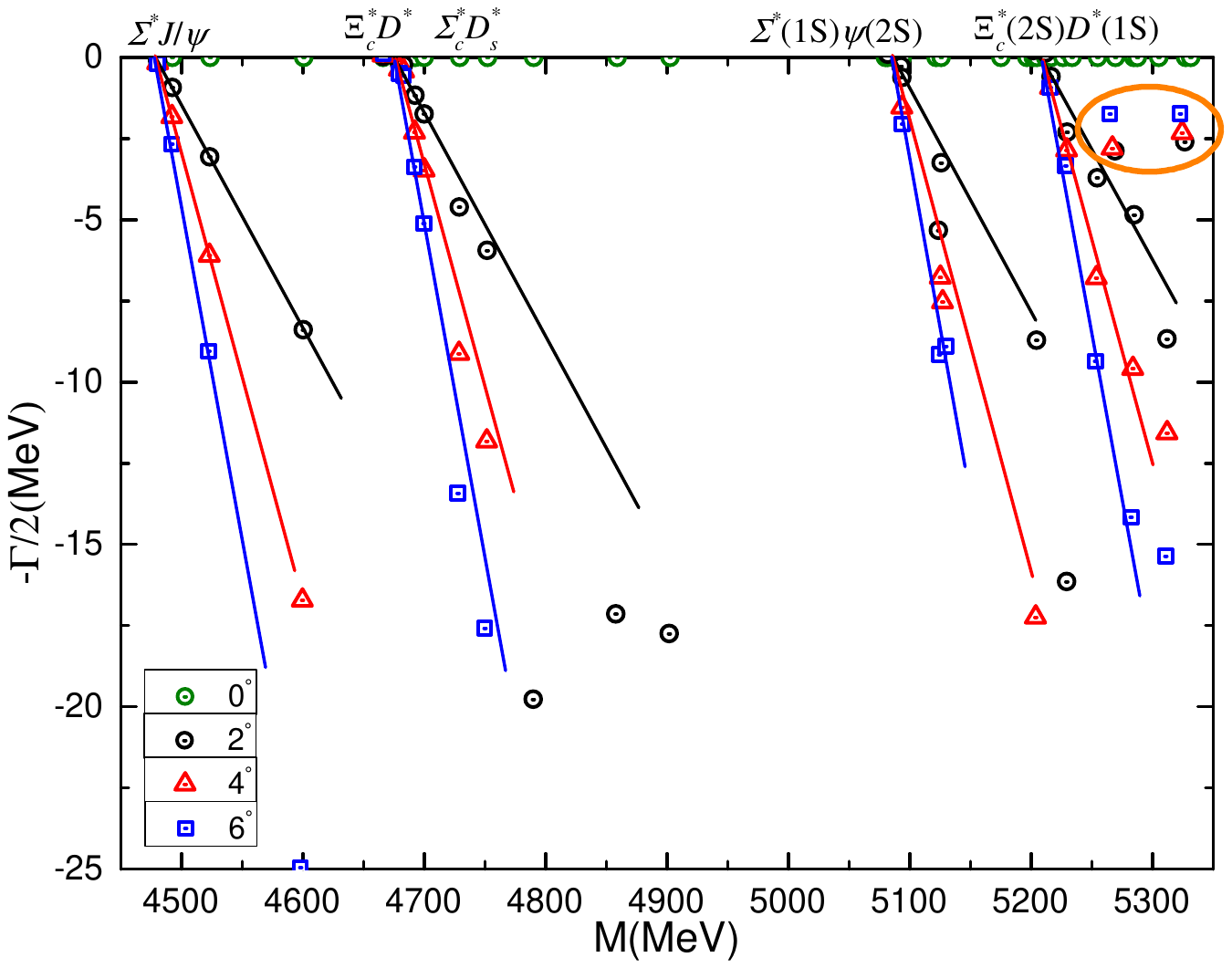}
\caption{ Complex energies of complete coupled-channel calculation for the $qqsc\bar{c}$ pentaquark within $IJ^P=1(\frac52^-)$.} \label{PP6}
\end{figure}

The CSM is employed in a fully coupled-channel calculation, we show in Fig.~\ref{PP1} to~\ref{PP6} the distribution of complex eigenenergies and, therein, the obtained bound and resonance states are indicated inside colored circles. Furthermore, within a real-range calculation of the complete coupled-channel case, some insights on the nature of exotic states are given by computing their sizes and probabilities of the different pentaquark configurations in their wave functions, the results are listed among Tables~\ref{GresultR1},~\ref{GresultR2},~\ref{GresultR3},~\ref{GresultR4} and~\ref{GresultR5}. Finally, a summary of our most salient results is presented in Table~\ref{GresultCCT}.

We proceed now to describe in detail our theoretical findings:

\begin{table}[!t]
\caption{\label{Gresult1} Lowest-lying $qqsc\bar{c}$ pentaquark states with $I(J^P)=0(\frac12^-)$ calculated in a real range formulation of the potential model.
Baryon-meson configuration is listed in the first column, the superscripts 1 and 8 stand for color-singlet and -octet states, respectively. The experimental threshold value of a baryon-meson channel is listed in the parenthesis. The lowest theoretical mass obtained in each channel is shown in the second column, and the binding energy is presented in the following one. A mixture of color-singlet and -octet state for each baryon-meson configuration is considered, the coupled mass and binding energy is shown in the last column.
The lowest-lying mass in a partial coupled-channel calculation, which includes the color-singlet, -octet channels coupling, and a complete coupled-channel one is performed, and they are indicated in the bottom of the table, respectively (unit: MeV).}
\begin{ruledtabular}
\begin{tabular}{lccc}
~~Channel  & $M$ & $E_B$ & Mixed ~~ \\
                  &  & & ($M_{1\oplus 8}$, $E_B$)~~ \\
\hline
$(\Lambda \eta_c)^1 (4097)$         & $3918$ & $0$ & $(3918, 0)$ \\
$(\Lambda \eta_c)^8$                 & $4782$  &$+864$ & \\ [2ex]
$(\Lambda_c D_s)^1 (4255)$         & $4029$ & $0$ & $(4029, 0)$  \\
$(\Lambda_c D_s)^8$                  & $4730$ & $+701$ &  \\ [2ex]
$(\Xi'_c D)^1 (4448)$                    & $4509$ & $-5$ & $(4058, -6)$ \\
$(\Xi'_c D)^8 $                   & $4818$ & $+304$ & \\ [2ex]
$(\Xi_c D)^1 (4340)$                     & $4289$ & $-11$ & $(4289, -11)$ \\
$(\Xi_c D)^8 $                    & $4762$ & $+462$ & \\ [2ex]
$(\Lambda J/\psi)^1 (4213)$         & $4025$ & $0$ & $(4025, 0)$ \\
$(\Lambda J/\psi)^8 $       & $4759$ & $+734$ & \\ [2ex]
$(\Lambda_c D^*_s)^1 (4399)$     & $4165$ & $0$ & $(4165, 0)$ \\
$(\Lambda_c D^*_s)^8 $    & $4677$ & $+512$ & \\ [2ex]
$(\Xi'_c D^*)^1 (4585)$               & $4627$ & $-4$ & $(4614, -17)$ \\
$(\Xi'_c D^*)^8$                & $4783$ & $+152$ & \\ [2ex]
$(\Xi_c D^*)^1 (4477)$                & $4410$ & $-7$ & $(4408, -9)$  \\
$(\Xi_c D^*)^8$                 & $4686$ & $+269$ &  \\ [2ex]
$(\Xi^*_c D^*)^1 (4652)$            & $4671$ & $-5$ & $(4617, -59)$ \\
$(\Xi^*_c D^*)^8$             & $4657$ & $-19$ &  \\ [2ex]
\multicolumn{3}{c}{All of color-singlet channels coupling:} & $3918$ \\
\multicolumn{3}{c}{All of color-octet channels coupling:} & $4653$ \\
\multicolumn{3}{c}{Complete coupled-channel:} & $3918$
\end{tabular}
\end{ruledtabular}
\end{table}

\begin{table}[!t]
\caption{\label{GresultR1} Compositeness of exotic resonances obtained in a complete coupled-channel computation in $0(\frac12^-)$ state of $qqsc\bar{c}$ pentaquark. Particularly, the first column is the resonance pole labeled by $M+i\Gamma$, unit in MeV; the second one is the distance between any two quarks $(q=u, d, s)$ or quark-antiquark, unit in fm; and the dominant component of resonance state ($S$: baryon-meson structure in color-singlet channel; $H$: baryon-meson structure in hidden-color channel).}
\begin{ruledtabular}
\begin{tabular}{lccc}
Resonance       & \multicolumn{3}{c}{Structure} \\[2ex]
$4303+i1.0$   & \multicolumn{3}{c}{$r_{qq}:0.99$;\,\,\,\,\,$r_{qc}:1.50$;\,\,\,\,\,$r_{q \bar{c}}:1.59$;\,\,\,\,$r_{c \bar{c}}:0.97$} \\ [2ex]
& \multicolumn{3}{c}{$S$: 90.7\%;\,\, $H$: 9.3\%} \\
& \multicolumn{3}{c}{$(\Lambda J/\psi)^1$: 60\%;\,\, $(\Xi_c D)^1$: 23\%} \\ [2ex]
$4603+i15.9$   & \multicolumn{3}{c}{$r_{qq}:1.06$;\,\,\,\,\,$r_{qc}:1.80$;\,\,\,\,\,$r_{q \bar{c}}:1.88$;\,\,\,\,$r_{c \bar{c}}:1.16$} \\ [2ex]
& \multicolumn{3}{c}{$S$: 89\%;\,\, $H$: 11\%} \\
& \multicolumn{3}{c}{$(\Lambda \eta_c)^1$: 21\%;\,\, $(\Lambda J/\psi)^1$: 29\%;\,\, $(\Lambda_c D_s)^1$: 13\%}
\end{tabular}
\end{ruledtabular}
\end{table}

{\bf The $\bm{I(J^P)=0(\frac12^-)}$ channel:} All of the possible baryon-meson channels, $\Lambda \eta_c$, $\Lambda J/\psi$, $\Lambda_c D^{(*)}_s$, $\Xi'_c D^{(*)}$ and $\Xi^{(*)}_c D^{(*)}$, listed in Table~\ref{Gresult1}, are firstly investigated in a real-range calculation. The lowest channel, $\Lambda \eta_c$, has a theoretical mass of $3918$ MeV, which is just the theoretical threshold value, and the it is a scattering state. The unbound nature also holds for other $(qqs)(c\bar{c})$ and $(qqc)(s\bar{c})$ configurations, \emph{viz.} color-singlet channels of $\Lambda J/\psi$ and $\Lambda_c D^{(*)}_s$ configurations are all of scattering type. Besides, the coupling effect is quite weak in these cases when considering their respective hidden-color channels; hence, the scattering nature is remained. On the other hand, bound states are found in the $(qsc)(q\bar{c})$ configuration; particularly, five baryon-meson channels contribute, \emph{i.e.} $\Xi'_c D$, $\Xi_c D$, $\Xi'_c D^*$, $\Xi_c D^*$ and $\Xi^*_c D^*$. There are binding energies which range from $-4$ to $-11$ MeV for the color-singlet channels. Concerning the $\Xi_c D$ channel, which has $-11$ MeV binding energy and then it is at $4.33$ GeV attending to its experimental threshold, it is quite compatible with the reported $P_{cs}(4338)$ state~\cite{LHCb:2022ogu}. However, the stability of this state needs to be confirmed in a further coupled-channel analysis. Meanwhile, hidden color channels of these kind of configurations predict unbounded states, except for a $-19$ MeV binding energy of the color-octet $\Xi^*_c D^*$. This color resonance gets more tightly bound with $M=4617$ MeV and $E_B=-59$ MeV if the singlet and octet channels are all coupled, while the coupling is weak in other $\Xi^{(*)}_c D^{(*)}$ channels.

In a further step, three types of coupled-channel computations: all color-singlets, all color-octets and fully-coupled, are performed with $\theta=0^\circ$ (real range calculation). The lowest-lying masses are listed in the bottom of Table~\ref{Gresult1}. Particularly, the scattering nature of the lowest channel, $\Lambda \eta_c$, remains in this kind of computations; moreover, a color-octet resonant signal at $4653$ MeV is also obtained.

In order to better understand the spectrum of $qqsc\bar{c}$ pentaquarks with quantum numbers $I(J^P)=0(\frac12^-)$, the CSM is adopted by considering a rotated angle ranging from $2^\circ$ to $6^\circ$. The distribution of calculated complex energies are plotted in Fig.~\ref{PP1}. Therein, with an energy interval from $3.9$ to $4.7$ GeV, the nine scattering states of $\Lambda \eta_c$, $\Lambda J/\psi$, $\Lambda_c D^{(*)}_s$, $\Xi'_c D^{(*)}$ and $\Xi^{(*)}_c D^{(*)}$ are well presented. The vast majority of energy dots are aligned along the corresponding threshold lines; however, two stable poles are obtained and they are circled.

Table~\ref{GresultR1} collects information about the two resonances obtained in a complete coupled-channel computation by the CSM. Firstly, their masses and widths $(M,\,\Gamma)$ are $(4303, 1.0)$ MeV and $(4603, 15.9)$ MeV, respectively. Apparently, the lower resonance can be identified as the $P_{cs}(4338)$ state. The dominant two-body strong decay widths are the color-singlet channels $\Lambda J/\psi (60\%)$ and $\Xi_c D (23\%)$. Its size is less than $1.6$ fm. Moreover, the higher resonance at $4.6$ GeV has a width of $15.9$ MeV, which is mainly given by $\Lambda \eta_c (21\%)$, $\Lambda J/\psi (29\%)$ and $\Lambda_c D_s (13\%)$ final states in the singlet color channel. This exotic state, whose size is less than $1.9$ fm, is expected to be confirmed in future experiments.


\begin{table}[!t]
\caption{\label{Gresult2} Lowest-lying $qqsc\bar{c}$ pentaquark states with $I(J^P)=0(\frac32^-)$ calculated in a real range formulation of the potential model. The table is similarly organized as Table~\ref{Gresult1}.
(unit: MeV).}
\begin{ruledtabular}
\begin{tabular}{lccc}
~~Channel  & $M$ & $E_B$ & Mixed ~~ \\
                  &  & & ($M_{1\oplus 8}$, $E_B$)~~ \\
\hline
$(\Lambda J/\psi)^1 (4213)$         & $4025$ & $0$ & $(4025, 0)$ \\
$(\Lambda J/\psi)^8$                 & $4793$  &$+768$ & \\ [2ex]
$(\Lambda_c D^*_s)^1 (4399)$         & $4166$ & $0$ & $(4166, 0)$  \\
$(\Lambda_c D^*_s)^8$                  & $4746$ & $+580$ &  \\ [2ex]
$(\Xi'_c D^*)^1 (4585)$                    & $4628$ & $-3$ & $(4628, -3)$ \\
$(\Xi'_c D^*)^8 $                   & $4898$ & $+267$ & \\ [2ex]
$(\Xi_c D^*)^1 (4477)$                     & $4411$ & $-7$ & $(4411, -7)$ \\
$(\Xi_c D^*)^8 $                    & $4787$ & $+369$ & \\ [2ex]
$(\Xi^*_c D)^1 (4515)$            & $4555$ & $-4$ & $(4553, -6)$ \\
$(\Xi^*_c D)^8$             & $4853$ & $+294$ &  \\ [2ex]
$(\Xi^*_c D^*)^1 (4652)$            & $4674$ & $-2$ & $(4656, -20)$ \\
$(\Xi^*_c D^*)^8$             & $4771$ & $+95$ &  \\ [2ex]
\multicolumn{3}{c}{All of color-singlet channels coupling:} & $4025$ \\
\multicolumn{3}{c}{All of color-octet channels coupling:} & $4725$ \\
\multicolumn{3}{c}{Complete coupled-channel:} & $4025$
\end{tabular}
\end{ruledtabular}
\end{table}

\begin{table}[!t]
\caption{\label{GresultR2} Compositeness of exotic resonances obtained in a complete coupled-channel computation in $0(\frac32^-)$ state of $qqsc\bar{c}$ pentaquark. Results are similarly organized as those in Table~\ref{GresultR1}.}
\begin{ruledtabular}
\begin{tabular}{lccc}
Resonance       & \multicolumn{3}{c}{Structure} \\[2ex]
$4419+i0.5$   & \multicolumn{3}{c}{$r_{qq}:1.80$;\,\,\,\,\,$r_{qc}:1.32$;\,\,\,\,\,$r_{q \bar{c}}:1.75$;\,\,\,\,$r_{c \bar{c}}:2.03$} \\ [2ex]
& \multicolumn{3}{c}{$S$: 89.6\%;\,\, $H$: 10.4\%} \\
& \multicolumn{3}{c}{$(\Xi_c D^*)^1$: 72\%} \\ [2ex]
$4659+i5.4$   & \multicolumn{3}{c}{$r_{qq}:1.70$;\,\,\,\,\,$r_{qc}:2.15$;\,\,\,\,\,$r_{q \bar{c}}:2.40$;\,\,\,\,$r_{c \bar{c}}:1.94$} \\ [2ex]
& \multicolumn{3}{c}{$S$: 93.8\%;\,\, $H$: 6.2\%} \\
& \multicolumn{3}{c}{$(\Lambda J/\psi)^1$: 28\%;\,\, $(\Xi'_c D^*)^1$: 19\%} \\
& \multicolumn{3}{c}{$(\Xi_c D^*)^1$: 19\%;\,\, $(\Xi^*_c D)^1$: 13\%}
\end{tabular}
\end{ruledtabular}
\end{table}

{\bf The $\bm{I(J^P)=0(\frac32^-)}$ channel:} Table~\ref{Gresult2} lists our results of hidden-charm pentaquarks with strangeness in the mentioned channel obtained by the real-range calculation. In particular, $\Lambda J/\psi$, $\Lambda_c D^*_s$, $\Xi'_c D^*$ and $\Xi^{(*)}_c D^{(*)}$ are all the configurations considered. First, the lowest mass $4025$ MeV is the theoretical threshold value of $\Lambda J/\psi$; hence, it is just a scattering state. Besides, the second energy level, which lies at $4166$ MeV, is the theoretical threshold of $\Lambda_c D^*_s$, and the unbound nature is also concluded. The scattering feature of $(qqs)(c\bar{c})$ and $(qqc)(s\bar{c})$ configurations is characteristic of the coupled-channel studies that consider either singlet- and color-octet channels. Particularly, the exciting energy of $(\Lambda J/\psi)^8$ and $(\Lambda_c D^*_s)^8$ is $768$ and $580$ MeV, respectively. However, as in the channel $I(J^P)=0(\frac12^-)$, bound states are found in the $(qsc)(q\bar{c})$ configuration. From Table~\ref{Gresult2}, one can find $\sim 4$ MeV binding energies for color-singlet channels $\Xi'_c D^*$, $\Xi_c D^*$, $\Xi^*_c D$ and $\Xi^*_c D^*$. Their hidden-color or color-octet channels are generally $300$ MeV higher than theoretical thresholds, except for the $(\Xi^*_c D^*)^8$ state with $95$ MeV exciting energy. Additionally, after a mixture of the singlet- and hidden-color channels, the lowest mass of $\Xi^*_c D$ and $\Xi^*_c D^*$ shifts to $4553$ and $4656$ MeV, respectively. However, $\Xi'_c D^*$ and $\Xi_c D^*$ remains at $4268$ and $4411$ MeV, respectively. Herein, the $\Xi_c D^*$ bound state, which has a binding energy of $-7$ MeV and a modified mass $4470$ MeV, is compatible with the $P_{cs}(4459)$ in $I(J^P)=0(\frac32^-)$ state~\cite{LHCb:2020jpq}.

At the bottom of Table~\ref{Gresult2} we show the lowest coupled mass in three types of real-range calculations. When all of color-singlet channels are considered, the lowest mass, $4025$ MeV, is still the theoretical threshold value of $\Lambda J/\psi$ channel. This weakly coupling effect remains in the complete coupled-channel calculation. Besides, a color resonance at $4725$ MeV is obtained in a computation with all of hidden-color channels included.

The spectrum of $qqsc\bar{c}$ pentaquarks with isospin and spin-parity $0(\frac32^-)$ is now investigated in a fully-coupled calculation with the help of CSM, see Fig~\ref{PP2}. Within an energy range $4.0-4.8$ GeV, scattering states of  $\Lambda J/\psi$, $\Lambda_c D^*_s$, $\Xi_c D^*$, $\Xi^*_c D$, $\Xi'_c D^*$ and $\Xi^*_c D^*$ are clearly shown. However, two stable poles are obtained and circled. Their complex energies read as $4419+i0.5$ MeV and $4659+i5.4$ MeV, respectively. Moreover, quark--(anti-)quark distances and dominant components of resonances are listed in Table~\ref{GresultR2}. The first resonance at $4.42$ GeV is quite compatible with the $P_{cs}(4459)$ state~\cite{LHCb:2020jpq}. Its size is less than $2.0$ fm and the golden channel is $\Xi_c D^* (72\%)$ in our calculation. Besides, since the calculated distance between $q$ and $\bar{c}$ is $2.4$ fm, a loosely resonant nature of the second state at $4.66$ GeV can be drawn. There is a strong coupling among the color-singlet channels $\Lambda J/\psi(28\%)$, $\Xi'_c D^*(19\%)$, $\Xi_c D^*(19\%)$ and $\Xi^*_c D(13\%)$. Accordingly, the narrow resonance, $4659+i5.4$ MeV, is also expected to be found in future high energy experimental facilities. 


\begin{table}[!t]
\caption{\label{Gresult3} Lowest-lying $qqsc\bar{c}$ pentaquark states with $I(J^P)=0(\frac52^-)$ calculated in a real range formulation of the potential model. The table is similarly organized as Table~\ref{Gresult1}.
(unit: MeV).}
\begin{ruledtabular}
\begin{tabular}{lccc}
~~Channel  & $M$ & $E_B$ & Mixed ~~ \\
                  &  & & ($M_{1\oplus 8}$, $E_B$)~~ \\
\hline
$(\Xi^*_c D^*)^1 (4652)$            & $4673$ & $-3$ & $(4673, -3)$ \\
$(\Xi^*_c D^*)^8$             & $5003$ & $+327$ &  \\ 
\end{tabular}
\end{ruledtabular}
\end{table}

\begin{table}[!t]
\caption{\label{GresultR3} Compositeness of exotic states obtained in a complete coupled-channel computation in $0(\frac52^-)$ state of $qqsc\bar{c}$ pentaquark. Results are similarly organized as those in Table~\ref{GresultR1}.}
\begin{ruledtabular}
\begin{tabular}{lccc}
Exotic state       & \multicolumn{3}{c}{Structure} \\[2ex]
$4673+i0$   & \multicolumn{3}{c}{$r_{qq}:1.86$;\,\,\,\,\,$r_{qc}:1.36$;\,\,\,\,\,$r_{q \bar{c}}:1.78$;\,\,\,\,$r_{c \bar{c}}:2.05$} \\ [2ex]
$(E_B=-3)$ & \multicolumn{3}{c}{$(\Xi^*_c D^*)^1$: 97.4\%;\,\, $(\Xi^*_c D^*)^8$: 2.6\%} \\ [2ex]
$5533+i3.4$   & \multicolumn{3}{c}{$r_{qq}:1.55$;\,\,\,\,\,$r_{qc}:1.18$;\,\,\,\,\,$r_{q \bar{c}}:1.28$;\,\,\,\,$r_{c \bar{c}}:0.89$} \\ [2ex]
& \multicolumn{3}{c}{$(\Xi^*_c D^*)^1$: 49.5\%;\,\, $(\Xi^*_c D^*)^8$: 50.5\%} \\
\end{tabular}
\end{ruledtabular}
\end{table}

{\bf The $\bm{I(J^P)=0(\frac52^-)}$ channel:} Only one baryon-meson channel, $\Xi^* D^*$, contributes to the highest spin channel within the isoscalar sector. Firstly, in the single channel calculation that includes the color-singlet and -octet configurations, the lowest-lying mass is $4673$ and $5003$ MeV, respectively, which corresponds to a binding energy of $-3$ and $327$ MeV, when comparing to the theoretical threshold. Moreover, the channel-coupling effect is extremely weak in this case and thus the coupled mass remains at $4673$ MeV.

A complex-range analysis of fully coupled-channel calculation is then performed, and results are presented in Fig.~\ref{PP3}. In the $4.6-5.6$ GeV energy region, three scattering states, which include the $\Xi^*(1S) D^*(1S)$ and its radial excited cases $\Xi^*(2S) D^*(1S)$ and $\Xi^*(1S) D^*(2S)$, are well presented. Moreover, one bound state and one narrow resonance are also obtained. Firstly, the mentioned $\Xi^* D^*$ appears again loosely bound. Secondly, a narrow resonance with $\Gamma=3.4$ MeV is obtained at $5533$ MeV. It is compact, with size around $1.2$ fm, and there is a strong coupling between the color-singlet $(50\%)$ and -octet $(50\%)$ channels of $\Xi^* D^*$.


\begin{table}[!t]
\caption{\label{Gresult4} Lowest-lying $qqsc\bar{c}$ pentaquark states with $I(J^P)=1(\frac12^-)$ calculated in a real range formulation of the potential model. The table is similarly organized as Table~\ref{Gresult1}.
(unit: MeV).}
\begin{ruledtabular}
\begin{tabular}{lccc}
~~Channel  & $M$ & $E_B$ & Mixed ~~ \\
                  &  & & ($M_{1\oplus 8}$, $E_B$)~~ \\
\hline
$(\Sigma \eta_c)^1 (4174)$         & $4084$ & $0$ & $(4084, 0)$ \\
$(\Sigma \eta_c)^8$                 & $4837$  &$+753$ & \\ [2ex]
$(\Sigma J/\psi)^1 (4290)$         & $4192$ & $0$ & $(4192, 0)$  \\
$(\Sigma J/\psi)^8$                  & $4818$ & $+626$ &  \\ [2ex]
$(\Sigma^* J/\psi)^1 (4482)$         & $4477$ & $0$ & $(4477, 0)$ \\
$(\Sigma^* J/\psi)^8 $                   & $4807$ & $+330$ & \\ [2ex]
$(\Sigma_c D_s)^1 (4422)$                     & $4501$ & $0$ & $(4501, 0)$ \\
$(\Sigma_c D_s)^8 $                    & $4906$ & $+405$ & \\ [2ex]
$(\Sigma_c D^*_s)^1 (4566)$            & $4636$ & $0$ & $(4636, 0)$ \\
$(\Sigma_c D^*_s)^8$             & $4877$ & $+241$ &  \\ [2ex]
$(\Sigma^*_c D^*_s)^1 (4632)$            & $4679$ & $0$ & $(4621, -58)$ \\
$(\Sigma^*_c D^*_s)^8$             & $4645$ & $-34$ &  \\ [2ex]
$(\Xi'_c D)^1 (4448)$            & $4514$ & $0$ & $(4514, 0)$ \\
$(\Xi'_c D)^8$             & $4909$ & $+395$ &  \\ [2ex]
$(\Xi_c D)^1 (4340)$            & $4301$ & $0$ & $(4301, 0)$ \\
$(\Xi_c D)^8$             & $4752$ & $+451$ &  \\ [2ex]
$(\Xi'_c D^*)^1 (4585)$            & $4631$ & $0$ & $(4631, 0)$ \\
$(\Xi'_c D^*)^8$             & $4908$ & $+277$ &  \\ [2ex]
$(\Xi_c D^*)^1 (4477)$            & $4418$ & $0$ & $(4418, 0)$ \\
$(\Xi_c D^*)^8$             & $4695$ & $+277$ &  \\ [2ex]
$(\Xi^*_c D^*)^1 (4652)$            & $4676$ & $0$ & $(4676, 0)$ \\
$(\Xi^*_c D^*)^8$             & $4857$ & $+181$ &  \\ [2ex]
\multicolumn{3}{c}{All of color-singlet channels coupling:} & $4084$ \\
\multicolumn{3}{c}{All of color-octet channels coupling:} & $4563$ \\
\multicolumn{3}{c}{Complete coupled-channel:} & $4084$
\end{tabular}
\end{ruledtabular}
\end{table}

{\bf The $\bm{I(J^P)=1(\frac12^-)}$ channel:} Table~\ref{Gresult4} lists real-range calculations of the $qqsc\bar{c}$ pentaquarks with iso-vector character and spin-parity $\frac12^-$. We consider 11 baryon-meson configurations and they are $\Sigma \eta_c$, $\Sigma^{(*)} J/\psi$, $\Sigma^{(*)}_c D^{(*)}_s$, $\Xi'_c D^{(*)}$ and $\Xi^{(*)}_c D^{(*)}$. Firstly, the lowest-lying state in color-singlet channels is $\Sigma \eta_c$ with calculated mass $4084$ MeV. Since its mass is located just at the theoretical threshold value, a scattering nature is deduced and this unbound feature also holds for other singlet color channels. However, one bound state with mass and binding energy $4645$ and $-34$ MeV, respectively, is found in the hidden-color channel of $\Sigma^*_c D^*_s$ state. When a color-structure mixture is considered, this bound state is pushed down towards $4621$ MeV, with a deeper binding energy of $-58$ MeV. Other hidden-color channels are generally $200-750$ MeV higher than their corresponding thresholds, and the coupled-channels mechanism of color structures does not help in forming a bound state.

Furthermore, as shown in the bottom of Table~\ref{Gresult4}, in the three types of coupled-channel calculations, the scattering nature of $\Sigma \eta_c$ state remains unchanged while a color resonance at $4563$ MeV is obtained by only considering hidden-color channels coupling.

The stability of the bound and color resonance states at $4621$ and $4563$ MeV, respectively, should be further studied in a complete coupled-channel calculation by using the CSM. Figure~\ref{PP4} shows the distribution of complex energies. Particularly, scattering states of $\Sigma \eta_c$, $\Sigma J/\psi$, $\Xi_c D$, $\Xi_c D^*$ and $\Sigma^* J/\psi$ are presented in the top panel, and no stable pole is found within $4.0-4.5$ GeV. An enlarged part from $4.49-4.59$ GeV is plotted in the middle panel of Fig~\ref{PP4}. Therein, resonance pole is still unavailable, and the scattering states of $\Sigma_c D_s$ and $\Xi'_c D$ are presented. Hence the previous color resonance at $4563$ MeV do not survive in a fully coupled-channel case. Finally, in the bottom panel, whose energy range goes from $4.62$ to $4.70$ GeV, four scattering states corresponding to $\Xi' D^*$, $\Sigma_c D^*_s$, $\Xi^*_c D^*$ and $\Sigma^*_c D^*_s$ are shown, and there is no evidence of a resonance state. Accordingly, the $(\Sigma^*_c D^*)^8$ bound state, which was obtained in a partial channels coupling computation, is quite unstable.


\begin{table}[!t]
\caption{\label{Gresult5} Lowest-lying $qqsc\bar{c}$ pentaquark states with $I(J^P)=1(\frac32^-)$ calculated in a real range formulation of the potential model. The table is similarly organized as Table~\ref{Gresult1}.
(unit: MeV).}
\begin{ruledtabular}
\begin{tabular}{lccc}
~~Channel  & $M$ & $E_B$ & Mixed ~~ \\
                  &  & & ($M_{1\oplus 8}$, $E_B$)~~ \\
\hline
$(\Sigma J/\psi)^1 (4290)$         & $4192$ & $0$ & $(4192, 0)$  \\
$(\Sigma J/\psi)^8$                  & $4845$ & $+653$ &  \\ [2ex]
$(\Sigma^* \eta_c)^1 (4366)$            & $4370$ & $0$ & $(4370, 0)$ \\
$(\Sigma^* \eta_c)^8$             & $4857$ & $+487$ &  \\ [2ex]
$(\Sigma^* J/\psi)^1 (4482)$         & $4477$ & $0$ & $(4477, 0)$ \\
$(\Sigma^* J/\psi)^8 $                   & $4837$ & $+360$ & \\ [2ex]
$(\Sigma_c D^*_s)^1 (4566)$            & $4636$ & $0$ & $(4636, 0)$ \\
$(\Sigma_c D^*_s)^8$             & $4912$ & $+276$ &  \\ [2ex]
$(\Sigma^*_c D_s)^1 (4488)$            & $4543$ & $0$ & $(4543, 0)$ \\
$(\Sigma^*_c D_s)^8$             & $4735$ & $+192$ &  \\ [2ex]
$(\Sigma^*_c D^*_s)^1 (4632)$            & $4679$ & $0$ & $(4669, -10)$ \\
$(\Sigma^*_c D^*_s)^8$             & $4698$ & $+19$ &  \\ [2ex]
$(\Xi'_c D^*)^1 (4585)$            & $4631$ & $0$ & $(4631, 0)$ \\
$(\Xi'_c D^*)^8$             & $4888$ & $+257$ &  \\ [2ex]
$(\Xi_c D^*)^1 (4477)$            & $4418$ & $0$ & $(4418, 0)$ \\
$(\Xi_c D^*)^8$             & $4772$ & $+354$ &  \\ [2ex]
$(\Xi^*_c D)^1 (4515)$            & $4559$ & $0$ & $(4676, 0)$ \\
$(\Xi^*_c D)^8$             & $4856$ & $+297$ &  \\ [2ex]
$(\Xi^*_c D^*)^1 (4652)$            & $4676$ & $0$ & $(4676, 0)$ \\
$(\Xi^*_c D^*)^8$             & $4848$ & $+172$ &  \\ [2ex]
\multicolumn{3}{c}{All of color-singlet channels coupling:} & $4192$ \\
\multicolumn{3}{c}{All of color-octet channels coupling:} & $4524$ \\
\multicolumn{3}{c}{Complete coupled-channel:} & $4192$
\end{tabular}
\end{ruledtabular}
\end{table}

\begin{table}[!t]
\caption{\label{GresultR4} Compositeness of exotic resonances obtained in a complete coupled-channel computation in $1(\frac32^-)$ state of $qqsc\bar{c}$ pentaquark. Results are similarly organized as those in Table~\ref{GresultR1}.}
\begin{ruledtabular}
\begin{tabular}{lccc}
Resonance       & \multicolumn{3}{c}{Structure} \\[2ex]
$4625+i4.0$   & \multicolumn{3}{c}{$r_{qq}:1.60$;\,\,\,\,\,$r_{qc}:1.19$;\,\,\,\,\,$r_{q \bar{c}}:1.53$;\,\,\,\,$r_{c \bar{c}}:1.72$} \\ [2ex]
& \multicolumn{3}{c}{$S$: 87\%;\,\, $H$: 13\%} \\
& \multicolumn{3}{c}{$(\Sigma_c D^*_s)^1$: 33\%;\,\, $(\Sigma^*_c D_s)^1$: 12\%;\,\, $(\Xi'_c D^*)^1$: 21\%} \\ [2ex]
$4803+i3.9$   & \multicolumn{3}{c}{$r_{qq}:1.39$;\,\,\,\,\,$r_{qc}:1.60$;\,\,\,\,\,$r_{q \bar{c}}:1.77$;\,\,\,\,$r_{c \bar{c}}:1.40$} \\ [2ex]
& \multicolumn{3}{c}{$S$: 90.6\%;\,\, $H$: 9.4\%} \\
& \multicolumn{3}{c}{$(\Sigma^* J/\psi)^1$: 27\%;\,\, $(\Xi^*_c D)^1$: 21\%;\,\, $(\Xi^*_c D^*)^1$: 14\%}
\end{tabular}
\end{ruledtabular}
\end{table}

{\bf The $\bm{I(J^P)=1(\frac32^-)}$ channel:} 10 baryon-meson configurations listed in Table~\ref{Gresult5} are investigated herein. Among the $\Sigma^{(*)} J/\psi$, $\Sigma^* \eta_c$, $\Sigma^{(*)}_c D^{(*)}_s$, $\Xi'_c D^*$ and $\Xi^{(*)}_c D^{(*)}$ channels both in color-singlet and hidden-color arrangements, the lowest-lying one is $\Sigma J/\psi$, and its mass is $4192$ MeV, which is just the theoretical value of the non-interacting baryon-meson threshold. Moreover, bound states are still not obtained in other channel calculations, and hidden-color channels are generally excited by an energy of $200-650$ MeV, except the color-octet channel of $\Sigma^*_c D^*_s$ which is $19$ MeV higher than its theoretical threshold. Furthermore, a weakly bound state, whose mass and binding energy are $4669$ and $-10$ MeV, respectively, is obtained in the $\Sigma^*_c D^*_s$ configuration when the singlet- and hidden-color channels are mixed.

When we perform a coupled-channel calculation within the real-range formalism taking into account all singlet channels, all octet channels and a fully coupling case, bound states are not obtained, the lowest-lying mass, $4192$ MeV, is the $\Sigma J/\psi$ theoretical threshold value, and a color-octet resonance located at $4524$ MeV is found.

Figure~\ref{PP5} shows the distribution of complex energies in a fully coupled-channels study using the CSM. Within the mass interval of $4.15-4.85$ GeV, scattering states of $\Sigma^{(*)} J/\psi$, $\Sigma^* \eta_c$, $\Sigma^{(*)}_c D^{(*)}_s$, $\Xi'_c D^*$ and $\Xi^{(*)}_c D^{(*)}$ are clearly found. Apart from them, two stable poles are circled in the complex energy plane. Their nature and structural information can be found in Table~\ref{GresultR4}. In particular, the lower resonance is at $4625$ MeV and the higher one is at $4803$ MeV. Their two-body strong decay widths are $4$ MeV. Besides, they have similar sizes, which are about $1.5$ fm. Color-singlet channels account for the dominant contribution to their wave functions: $\Sigma_c D^*_s (33\%)$, $\Sigma^*_c D_s (12\%)$ and $\Xi'_c D^* (21\%)$ for the lower resonance whereas $\Sigma^* J/\psi (27\%)$, $\Xi^*_c D (21\%)$ and $\Xi^*_c D^* (14\%)$ for the other one.


\begin{table}[!t]
\caption{\label{Gresult6} Lowest-lying $qqsc\bar{c}$ pentaquark states with $I(J^P)=1(\frac52^-)$ calculated in a real range formulation of the potential model. The table is similarly organized as Table~\ref{Gresult1}.
(unit: MeV).}
\begin{ruledtabular}
\begin{tabular}{lccc}
~~Channel  & $M$ & $E_B$ & Mixed ~~ \\
                  &  & & ($M_{1\oplus 8}$, $E_B$)~~ \\
\hline
$(\Sigma^* J/\psi)^1 (4482)$         & $4477$ & $0$ & $(4477, 0)$ \\
$(\Sigma^* J/\psi)^8 $                   & $4883$ & $+406$ & \\ [2ex]
$(\Sigma^*_c D^*_s)^1 (4632)$     & $4679$ & $0$ & $(4679, 0)$ \\
$(\Sigma^*_c D^*_s)^8$             & $4773$ & $+94$ &  \\ [2ex]
$(\Xi^*_c D^*)^1 (4652)$            & $4676$ & $0$ & $(4676, 0)$ \\
$(\Xi^*_c D^*)^8$             & $4853$ & $+177$ &  \\ [2ex]
\multicolumn{3}{c}{All of color-singlet channels coupling:} & $4477$ \\
\multicolumn{3}{c}{All of color-octet channels coupling:} & $4700$ \\
\multicolumn{3}{c}{Complete coupled-channel:} & $4477$
\end{tabular}
\end{ruledtabular}
\end{table}

\begin{table}[!t]
\caption{\label{GresultR5} Compositeness of exotic states obtained in a complete coupled-channel computation in $1(\frac52^-)$ state of $qqsc\bar{c}$ pentaquark. Results are similarly organized as those in Table~\ref{GresultR1}.}
\begin{ruledtabular}
\begin{tabular}{lccc}
Resonance       & \multicolumn{3}{c}{Structure} \\[2ex]
$5269+i5.8$   & \multicolumn{3}{c}{$r_{qq}:1.39$;\,\,\,\,\,$r_{qc}:1.29$;\,\,\,\,\,$r_{q \bar{c}}:1.38$;\,\,\,\,$r_{c \bar{c}}:1.11$} \\ [2ex]
& \multicolumn{3}{c}{$S$: 54\%;\,\, $H$: 46\%} \\
& \multicolumn{3}{c}{$(\Sigma^* J/\psi)^1$: 43\%;\,\, $(\Sigma^*_c D^*_s)^8$: 29.4\%} \\ [2ex]
$5327+i5.2$   & \multicolumn{3}{c}{$r_{qq}:1.38$;\,\,\,\,\,$r_{qc}:1.19$;\,\,\,\,\,$r_{q \bar{c}}:1.28$;\,\,\,\,$r_{c \bar{c}}:1.11$} \\ [2ex]
& \multicolumn{3}{c}{$S$: 54.3\%;\,\, $H$: 45.7\%} \\
& \multicolumn{3}{c}{$(\Sigma^* J/\psi)^1$: 36.2\%;\,\, $(\Sigma^*_c D^*_s)^8$: 34.1\%}
\end{tabular}
\end{ruledtabular}
\end{table}

{\bf $\bm{I(J^P)=1(\frac52^-)}$ channel:} Three baryon-meson configurations should be considered in the highest spin case of the isovector sector, and they are indicated in Table~\ref{Gresult6}. Particularly, the lowest-lying state is $\Sigma^* J/\psi$ with a theoretical mass $4477$ MeV; the two others lie at $4679$ and $4676$ MeV for $\Sigma^*_c D^*_s$ and $\Xi^*_c D^*$, respectively. Hidden-color channels are at least $90$ MeV higher in energy than theoretical threshold lines. Accordingly, no bound states are found and this result is also obtained within coupled-channel calculations, see the bottom part of Table~\ref{Gresult6}. However, a color resonance located at $4.7$ GeV is obtained within a coupled-channels analysis in which only hidden-color configurations are included.

Additionally, when a complex-range investigation is performed, considering all of the $1\frac52^-$ channels, two narrow resonances are found. Figure~\ref{PP6} shows the scattering states corresponding to $\Sigma^* J/\psi$, $\Sigma^*_c D^*_s$ and $\Xi^*_c D^*$ within an energy region of $4.45-5.35$ GeV. Moreover, two stable poles are circled, their complex energies read $5269+i5.8$ and $5327+i5.2$ MeV, respectively. By looking at Table~\ref{GresultR5}, which provides structural information of the two singularities, a strong coupling effect between the color-singlet $(\sim54\%)$ and -octet $(\sim46\%)$ channels if found in both cases. Moreover, the dominant components are also the same: $(\Sigma^* J/\psi)^1$ and $(\Sigma^*_c D^*_s)^8$. Additionally, one can find also similarities between the two resonances when looking at their inner quark distances. Generally, their sizes are around $1.3$ fm.


\begin{table*}[!t]
\caption{\label{GresultCCT} Summary of exotic structures found in the $qqsc\bar{c}$ $(q=u,\,d)$ pentaquark systems. The first column shows the isospin, total spin and parity of each singularity. If available, the second column lists well known experimental states, which may be identified in our theoretical framework. The third column refers to the dominant configuration components, particularly, the superscripts 1 and 8 stand for color-singlet and -octet states, respectively. For a concise purpose, the component without superscripts is of singlet color state. Theoretical bound and resonance states are presented with the following notation: $(M, E_B)$ and $M+i\Gamma$ in the last column, respectively (unit: MeV).}
\begin{ruledtabular}
\begin{tabular}{lccc}
~ $I(J^P)$ & Experimental state & Dominant Component   & Theoretical pole~~ \\
\hline
~~$0(\frac12^-)$ & $P_{cs}(4338)$  & $\Lambda J/\psi (60\%)+\Xi_c D (23\%)$   & $4303+i1.0$~~  \\
    &  & $\Lambda \eta_c (21\%)+\Lambda J/\psi (29\%)+\Lambda_c D_s (13\%)$   & $4603+i15.9$~~ \\[2ex]
~~$0(\frac32^-)$ & $P_{cs}(4459)$  & $\Xi_c D^* (72\%)$   & $4419+i0.5$~~  \\
    &  & $\Lambda J/\psi (28\%)+\Xi^{(')}_c D^* (19\%)+\Xi^*_c D (13\%)$   & $4659+i5.4$~~ \\[2ex]
~~$0(\frac52^-)$ &  & $\Xi^*_c D^* (97.4\%)$   & $(4673, -3)$~~  \\
    &  & $(\Xi^*_c D^*)^1 (49.5\%)+(\Xi^*_c D^*)^8 (50.5\%)$   & $5533+i3.4$~~ \\[2ex]
~~$1(\frac32^-)$ &  & $\Sigma_c D^*_s (33\%)+\Sigma^*_c D_s (12\%)+\Xi'_c D^* (21\%)$   & $4625+i4.0$~~  \\
    &  & $\Sigma^* J/\psi (27\%)+\Xi^*_c D (21\%)+\Xi^*_c D^* (14\%)$   & $4803+i3.9$~~ \\[2ex]
~~$1(\frac52^-)$ &  & $(\Sigma^* J/\psi)^1 (43\%)+(\Sigma^*_c D^*_s)^8 (29.4\%)$   & $5269+i5.8$~~  \\
    &  & $(\Sigma^* J/\psi)^1 (36.2\%)+(\Sigma^*_c D^*_s)^8 (34.1\%)$   & $5327+i5.2$~~ \\
\end{tabular}
\end{ruledtabular}
\end{table*}

\section{SUMMARY}
\label{sec:summary}

The $S$-wave hidden-charm pentaquarks with strangeness, $qqsc\bar{c}$ $(q=u,\,d)$, whose spin-parity are $J^P=\frac12^-$, $\frac32^-$ and $\frac52^-$, and isospin either $0$ or $1$, have been systematically investigated within a chiral quark model approach that employs a highly accurate computational method, the Gaussian expansion formalism (GEM), along with the complex scaling technique (CSM), which is a powerful tool when dealing simultaneously with bound, resonant and scattering states.

Within this theoretical framework, and by considering baryon-meson configurations in both singlet- and hidden-color channels, the two experimentally reported $P_{cs}$ states~\cite{LHCb:2020jpq, LHCb:2022ogu} can be well identified. Besides, other structures can be distinguished in the different channels, except for the $I(J^P)=1(\frac12^-)$ one. Table~\ref{GresultCCT} summarizes our theoretical findings on $qqsc\bar{c}$ pentaquarks. In particular, the $I(J^P)$ quantum numbers are indicated in the first column, plausible experimental assignments are listed in the following one, the third column shows the dominant components in the wavefunction of exotic states, and their theoretical pole positions are presented in the last column. 

The following details of our analysis are of particular interest. Firstly, the experimentally reported $P_{cs}(4338)$ and $P_{cs}(4459)$ signals, whose tentative assignments of spin-parity are $\frac12^-$ and $\frac32^-$, respectively, can be well identified within our theoretical framework as molecules in the isoscalar sector. The dominant components of the lower-energy state are $\Lambda J/\psi$ $(60\%)$ and $\Xi_c D$ $(23\%)$, while it is the $\Xi_c D^*$ $(72\%)$ structure that is dominant in the higher-energy candidate.
Secondly, narrow resonances are obtained in all of the allowed $I(J^P)$-channels, except for the $1(\frac12^-)$ one. Generally, they are located in a mass region from $4.6$ to $5.5$ GeV, and they have strong couplings to different color-singlet channels. In the $J^P=\frac52^-$ channels, both isoscalar and isovector, the color-singlet and -octet configurations couple strongly.
Finally, a $\Xi^*_c D^*$ shallow bound state is obtained in the $0(\frac52^-)$ channel. The theoretical mass and binding energy is $4673\,\text{MeV}$ and $-3\,\text{MeV}$, respectively. 

All of the above findings are expected to be confirmed in future high energy experiments.


\begin{acknowledgments}
Work partially financed by National Natural Science Foundation of China under Grant Nos. 12305093, 11535005 and 11775118; Zhejiang Provincial Natural Science Foundation under Grant No. LQ22A050004; Ministerio Espa\~nol de Ciencia e Innovaci\'on under grant Nos. PID2019-107844GB-C22 and PID2022-140440NB-C22; the Junta de Andaluc\'ia under contract Nos. Operativo FEDER Andaluc\'ia 2014-2020 UHU-1264517, P18-FR-5057 and also PAIDI FQM-370.
\end{acknowledgments}


\bibliography{PcsPentaquarks}

\end{document}